\documentclass[aps,prd,twocolumn,amsmath,amssymb,superscriptaddress,nofootinbib,showpacs,preprintnumbers]{revtex4-1}

\usepackage{graphicx}
\usepackage{bm}
\usepackage{amsmath,amssymb}
\usepackage{aas_macros}
\usepackage{natbib}
\usepackage[colorlinks=true]{hyperref}

\begin{document}

%\preprint{APS/123-QED}

\title{Novel approaches to dark-matter detection using space-time separated clocks}% Force line breaks with \\

\author{Etienne Savalle}
\affiliation{SYRTE, Observatoire de Paris, Universit\'e PSL, CNRS, Sorbonne Universit\'e, LNE, 75014 Paris, France}
\author{Benjamin M. Roberts}
\affiliation{SYRTE, Observatoire de Paris, Universit\'e PSL, CNRS, Sorbonne Universit\'e, LNE, 75014 Paris, France}
\author{Florian Frank}
\affiliation{SYRTE, Observatoire de Paris, Universit\'e PSL, CNRS, Sorbonne Universit\'e, LNE, 75014 Paris, France}
\author{Paul-Eric Pottie}
\affiliation{SYRTE, Observatoire de Paris, Universit\'e PSL, CNRS, Sorbonne Universit\'e, LNE, 75014 Paris, France}
\author{Ben T. McAllister}
\affiliation{ARC Centre of Excellence for Engineered Quantum Systems, School of Physics, University of Western Australia, Crawley WA 6009, Australia}
\author{Conner Dailey}
\affiliation{Department of Physics, University of Nevada, Reno 89557, USA}
\author{Andrei Derevianko}
\affiliation{Department of Physics, University of Nevada, Reno 89557, USA}
\author{Peter Wolf}
\email{peter.wolf@obspm.fr}
\affiliation{SYRTE, Observatoire de Paris, Universit\'e PSL, CNRS, Sorbonne Universit\'e, LNE, 75014 Paris, France}
\email{peter.wolf@obspm.fr}

\date{\today}% It is always \today, today,

\begin{abstract}
We discuss the theoretical analysis and interpretation of space-time separated clock experiments in the context of a space-time varying scalar field that is non-universally coupled to the standard model fields. If massive, such a field is a candidate for dark matter and could be detected in laboratory experiments. We show that space-time separated experiments have the potential to probe a fundamentally different parameter space from more common co-located experiments, allowing decorrelation of previously necessarily correlated parameters. Finally, we describe such a space-time separated clock experiment currently running at the Paris Observatory, and present some preliminary results as a proof of principle. We use those results to estimate the potential reach of the experiment in dark matter searches.

%\begin{description}
%\end{description}
\end{abstract}

\pacs{Valid PACS appear here}% PACS, the Physics and Astronomy
\maketitle

%%%%%%%%%%%%%%%%%%%%%
\section{Introduction}
The nature of dark matter (DM) is one of the most important outstanding problems in physics today. Despite composing the majority of the matter in the universe, evidence for dark matter particles in direct detection experiments remains elusive \cite{Bertone2018}. So far, much of the focus has been on weakly-interacting massive particles (WIMPs) with GeV to TeV masses, but the lack of evidence for their existence is contributing to an increase in interest for more varied candidate models \cite{AtomicReview2017}.

One example is the recent surge of theoretical and experimental work on the possibility of ultra-light (typically $\ll$~eV) dark matter detection
using the outstanding accuracy achieved in atomic clocks, and more generally in time and frequency metrology \cite{DereviankoDM2014,Arvanitaki2014,StadnikDMalpha2015,Tilburg2015,Hees2016,
Wcislo2016,GPSDM2017,Hees2018,Wcislo2018,RobertsAsymm2018,GPSDM2018,Alonso2018,Wolf2018}. Most of that work is using a DM model where the DM is a massive scalar field that is non-universally coupled to the standard model (SM) fields. Such non-universal couplings lead to an apparent violation of the equivalence principle, which can be searched for either by free fall tests (tests of the weak equivalence principle) or by comparing clocks of different types and/or at different locations in space-time (tests of local position invariance). For a recent review of both types of experiments in this context see \cite{Hees2018}. 

For small masses ($\ll$~eV) occupation numbers in galactic halos are very high, and the scalar field can be described classically either as coherent oscillations or by macroscopic features such as topological defects. Here we focus on oscillating massive scalar fields as DM candidates. Most experimental work in this domain has explored the very low mass region ($\leq 10^{-14}$~eV) \cite{Tilburg2015,Hees2016,
Wcislo2016,GPSDM2017,Wcislo2018} owing to the fact that atomic clocks have typical measurement rates of no more than about 1~Hz, which is equivalent to a $\simeq 10^{-14}$~eV field oscillating at its Compton-De Broglie frequency ($\omega_m = m c^2/\hbar$). Furthermore, most theoretical and experimental works have investigated experiments involving clocks that are co-located in space time, as such set-ups are the most common and most accurate, and also because their theoretical analysis avoids complications related to the modelling of the evolution of their positions and of the clock comparison methods used (exceptions being \cite{GPSDM2017,Arvanitaki2018}).

We address both of these issues, by presenting an experiment that allows much higher measurement rates and amounts to comparing the same oscillator at different times. We provide a complete theoretical model of that experiment in the framework of an oscillating scalar field that is non-universally coupled to the SM. In doing so, we find that in the common interpretation of such a scalar field as a ``variation of fundamental constants" the experiment leads to a measurement of the variation of a dimensional constant (the electron mass $m_e$) which as such is not meaningful as it depends on the system of units used \cite{Uzan2003}. But, this is only the case in that particular interpretation, the experiment being perfectly meaningful within the more fundamental scalar field theory. For co-located clock experiments such issues do not arise, as the dependence on the system of units drops out in the differential measurement \cite{KozlovBudker2018}, but this is no longer the case when the clocks are separated in space-time.   We finally show some first preliminary results from such an experiment that is currently running at the Paris Observatory, and discuss its future prospects and potential reach in DM searches.

\section{Dark matter in the form of non-universally coupled scalar fields}\label{sec:DM_gen}

The theory of light scalar fields that are non-universally coupled to the SM (and thus violate the equivalence principle) has been developed in, e.g. refs. \cite{Damour2010,StadnikDMalpha2015,Arvanitaki2014}, with different couplings (linear or quadratic) and differing conventions and notations for the scalar field and coupling constants (for a recent exhaustive overview see \cite{Hees2018}). Here we choose the linear coupling model for simplicity, but all our conclusions also apply to the quadratic case.

We start from the action
\begin{align}
	S&=\frac{1}{c}\int d^4x \frac{\sqrt{-g}}{2\kappa}\left[R-2g^{\mu\nu}\partial_\mu\varphi\partial_\nu\varphi-V(\varphi)\right] \nonumber\\
	&+ \frac{1}{c}\int d^4x \sqrt{-g}\Bigg[ \mathcal L_\textrm{SM}[g_{\mu\nu},\Psi_i]  + \mathcal L_\textrm{int}[g_{\mu\nu},\varphi,\Psi_i]  \Bigg]\, , \label{eq:action}
\end{align}
where $\kappa=8\pi G/c^4$, $R$ is the Ricci scalar of the space-time metric $g_{\mu\nu}$, $\varphi$ is a dimensionless scalar-field, $\mathcal L_\textrm{SM}$ is the Lagrangian density of the Standard Model of particles depending on the standard model fields $\Psi_i$, and $\mathcal L_\textrm{int}$ parametrizes the interaction between matter and the scalar field. We consider the linear coupling case
\begin{align}\label{eq:Lint_lin}
 \mathcal L_\textrm{int}&=	\varphi  \Bigg[d_e\frac{e^2 c}{16\pi\hbar\alpha}F^2-d_g\frac{\beta_3}{2g_3}\left(F^A\right)^2\\
& \quad -c^2\sum_{i=e,u,d}\Big(d_{m_i}+\gamma_{m_i}d_g\Big)m_i\bar\psi_i\psi_i\Bigg] \, , \nonumber
\end{align}
with $F_{\mu\nu}$ the standard electromagnetic Faraday tensor, $e$ the electron charge, $\alpha$ the fine structure constant, $F^A_{\mu\nu}$ the gluon strength tensor, $g_3$ the QCD gauge coupling, $\beta_3$ the $\beta$ function for the running of $g_3$, $m_i$ the mass of the fermions (electron and light quarks \footnote{Following the more recent literature \cite{dzuba:2008uq}, we do not take into account the effects of the strange quark, although they have been estimated in the past for atomic clock measurements \cite{Flambaum2004,Flambaum2006}.}, $\gamma_{m_i}$ the anomalous dimension giving the energy running of the masses of the QCD coupled fermions and $\psi_i$ the fermion spinors.  The constants $d_j$ characterize the interaction between the scalar field $\varphi$ and the different SM sectors.

Introducing a quadratic potential,
\begin{equation}
	V(\varphi)=2\frac{c^2}{\hbar^2}m^2\varphi^2\, ,
\end{equation}
leads to an oscillating solution for the scalar field (see \cite{Hees2016,Hees2018} for details) of the form
\begin{align}\label{eq:phi_1}
	\varphi(t,\bm x) &=  \varphi_0 \cos \left(\omega t - \bm k .\bm x +\delta\right), 
\end{align}
where  $\left|\bm k\right|^2 + c^2 m^2/\hbar^2=\omega^2/c^2$.

If the scalar field $\varphi$ is responsible for the DM in our galaxy, then $\bm k$ is given by the DM velocity distribution in the solar system, with typically $\hbar k/m \approx 10^{-3} c$ so that the $\bm k .\bm x$ term is negligible for the experiments discussed here \footnote{Due to the velocity distribution of DM in the galaxy, the $\bm k .\bm x$ term leads to a limitation of the coherence time of the oscillations to about $10^6$ periods. This is neglected here, but will be taken into account in the final analysis of our experimental results.}. Furthermore, the amplitude $\varphi_0$ is determined by the local DM energy density ($\rho \approx 0.4$~GeV/cm$^3$ \cite{mcmillan:2011vn}) as
\begin{equation}\label{eq:phi_0}
\varphi_0 = \left(\frac{\hbar^2\kappa\,\rho}{m^2c^2}\right)^{1/2}.
\end{equation}

\subsection{Interpretation in terms of varying fundamental constants}\label{sec:fund_const}
When comparing the interaction Lagrangian density (\ref{eq:Lint_lin}) to the SM part
\begin{align}\label{eq:L_SM}
 \mathcal L_\textrm{SM}&=-\frac{e^2 c}{16\pi\hbar\alpha}F^2-\frac{\beta_3}{2g_3}\left(F^A\right)^2\\
& \quad -c^2\sum_{i=e,u,d}\Big(1+\gamma_{m_i}\Big)m_i\bar\psi_i\psi_i \, , \nonumber
\end{align}
one can see directly that the coupling constants $d_i$ and the field $\varphi$ can be simply interpreted as a rescaling of five fundamental constants
\begin{subequations}\label{eq:constants}
	\begin{align}
		\alpha(\varphi)&= \alpha\left(1+d_e\varphi\right) \, , \\
		m_i(\varphi)&=m_i\left(1+d_{m_i}\varphi\right) \quad \textrm{for } i=e,u,d\,  \\
		\Lambda_3(\varphi)&= \Lambda_3\left(1+d_{g}\varphi\right) \, ,
	\end{align}
\end{subequations}
turning them into space-time varying quantities through their dependence on the field $\varphi(t, \bm x)$. The last identity involving the QCD mass scale $\Lambda_3$ is less straightforward than the other four, and derived in detail in \cite{Damour2010}. Also, in general, the quark masses are reduced to the average mass $m_q=(m_u+m_d)/2$ with \cite{Damour2010}
\begin{equation}
	m_q(\varphi)=m_q(1+d_{m_q}\varphi)\, , \textrm{ and } d_{m_q}=\frac{d_{m_u}m_u+d_{m_d}m_d}{m_u+m_d}\, .
\end{equation}

The point we want to make here, is that the interpretation in terms of varying constants ($\alpha, m_i, \Lambda_3$) is only that, i.e., a convenient interpretation. More specifically, equations (\ref{eq:constants}) only take that form when working in S.I. units and are therefore dependent on the system of units used.

To see this in a concrete example let us concentrate on only the electromagnetic and electron part of the Lagrangian density. Transforming both (\ref{eq:Lint_lin}) and (\ref{eq:L_SM}) to atomic units ($\hbar=e=m_e=4\pi\epsilon_0=1, c=1/\alpha$) gives
\begin{align}\label{eq:L_au}
 \mathcal L_\textrm{int}^\textrm{a.u.}&=\frac{d_e}{16\pi\alpha^2}F^2-\frac{d_{m_e}}{\alpha^2}\bar\psi_e\psi_e \\
 \mathcal L_\textrm{SM}^\textrm{a.u.}&=\frac{1}{16\pi\alpha^2}F^2-\frac{1}{\alpha^2}\bar\psi_e\psi_e
\end{align}
and now the correspondence between the couplings to the scalar field and the fundamental constants is less obvious.

This is of course straightforward and in no way changes the physics stemming from  non-universally coupled scalar fields, which is the same whichever system of units one uses. We simply want to point out, that what matters in that context is not ``which constants vary", but which sector of the SM Lagrangian is coupled to $\varphi$ by the coupling constants $d_i$. The one to one correspondence (\ref{eq:constants}) between fundamental constants and the $d_i$ depends on the system of units used. That simple correspondence is a very useful tool when analysing experiments (as we will see below) but should not be taken as more than that.

\section{Co-located and space-time separated clocks} \label{sec:clocks}

Quite generally the frequency of a gross structure (optical) atomic transition ``A" can be written as
\begin{equation}\label{eq:atom}
\nu_A = C_A \frac{\alpha^2 m_e c^2}{\hbar}F_A(\alpha)\, ,
\end{equation}
where $C_A$ is a numerical constant specific to transition A and $F_A$ is a dimensionless function of $\alpha$ also specific to the transition. Similarly the frequency of a solid resonant cavity ``C" can be written as
\begin{equation}\label{eq:cavity}
\nu_C = C_C \frac{\alpha m_e c^2}{\hbar}F_C(\alpha, m_e, m_q, \Lambda_3)\, .
\end{equation}
In the former case the $\alpha^2 m_e$ dependence comes from the Rydberg constant that determines the energy Eigenstates, in the latter case the $\alpha m_e$ dependence comes from the Bohr radius that determines the length of the solid. Additional dependencies may come from the functions $F_i$ and can be significant \cite{KozlovBudker2018,BetheBook,Stadnik2015b,Pasteka2018}. But for the arguments of this subsection they are not necessary and will be neglected.

The dependency of a particular frequency on fundamental constants can be parametrized in terms of sensitivity coefficients for each relevant constant $X$, defined as
\begin{equation}\label{equ:def_K}
\frac{\delta\nu}{\nu_0} = K_X\frac{\delta X}{X}.
\end{equation}
It is known, however, that the $K_X$ sensitivity coefficients actually depend on the system of units employed \cite{KozlovBudker2018}. This is easily seen, e.g., for the atomic transition or the cavity, which directly gives
\begin{equation}\label{eq:Kalpha_SI}
K_{A,\alpha}^{S.I.} = 2 \hspace{1 cm} K_{C,\alpha}^{S.I.} = 1\, .
\end{equation}
%and
%\begin{equation}\label{eq:Kme_SI}
%K_{A,m_e}^{S.I.} = 1 \hspace{1 cm} K_{C,m_e}^{S.I.} = 1\, .
%\end{equation}
However, when transforming both (\ref{eq:atom}) and (\ref{eq:cavity}) to atomic units we find
\begin{equation}\label{eq:Kalpha_au}
K_{A,\alpha}^{a.u.} = 0 \hspace{1 cm} K_{C,\alpha}^{a.u.} = -1\, .
\end{equation}
%and
%\begin{equation}\label{eq:Kme_au}
%K_{A,m_e}^{a.u.} = 0 \hspace{1 cm} K_{C,m_e}^{a.u.} = 0\, .
%\end{equation}

\subsection{Co-located clocks}
Nonetheless, meaningful (i.e., independent of the system of units used) experiments that search for a variation of fundamental constants can be conducting by comparing different types of clocks that are co-located. For example, the variation of the frequency ratio $\nu_A/\nu_C$   
\begin{equation}\label{eq:dnu_nu_X}
\frac{\delta(\nu_A/\nu_C)}{\nu_{A0}/\nu_{C0}} = (K_{A,X} - K_{C,X})\frac{\delta X}{X_0}\, ,
\end{equation}   
which when substituting (\ref{eq:Kalpha_SI}) {\it or} (\ref{eq:Kalpha_au}) provides a measurement of the possible variation of $\alpha$ 
\begin{equation}\label{eq:dnu_nu}
\frac{\delta(\nu_A/\nu_C)}{(\nu_{A}/\nu_{C})_0} = \frac{\delta \alpha}{\alpha_0}\, ,
\end{equation}
irrespective of the system of units used.

In terms of the underlying scalar field theory the effect on the experiment can be obtained directly by applying equations (\ref{eq:constants}) to (\ref{eq:dnu_nu}) giving
\begin{equation}\label{eq:dnu_nu_de}
\frac{(\nu_A/\nu_C)}{(\nu_{A}/\nu_{C})_0}(t, \bm x) = d_e\varphi(t,\bm x)\, .
\end{equation}

Note that any such co-located clock experiment can only provide a meaningful (i.e., independent of the system of units) result for dimensionless combination of fundamental constants, typically some combination of $\alpha$ and $m_i/\Lambda_3$, and correspondingly of $d_e$ and the {\bf difference} $d_{m_i}-d_g$, but not of the $d_{m_i}$ alone. 

\subsection{Space-time separated clocks} \label{sec:ST-sep_clocks}
Consider now an experiment where two clocks of the same type are separated into two regions of space-time where we suspect that the fundamental constants have different values. Is there an experiment we can do to determine if this is the case?

The two clocks are compared using light signals. In general, a variation of fundamental constants will also effect the light signals propagating through a fibre, and thus the link between the clocks, and that effect needs to be taken into account. But for the arguments of this section we will assume that the link is unaffected (this could be the case, e.g., in two-way links). We will include a full model of the fibre link when analysing the actual experiment in section \ref{sec:DAMNED}.

Atomic clock A2 is in a region where the fundamental constants have their nominal values $X_0$ and clock A1 in a region where they differ by $\delta X$. Then we can see directly from (\ref{eq:dnu_nu_X}) that
\begin{equation}\label{dnu_nu_12}
\frac{\delta(\nu_{A1}/\nu_{A2})}{(\nu_{A1}/\nu_{A2})_0} =  K_{A,X}\frac{\delta X}{X_0}\, .
\end{equation} 
Now, $K_{A,X}$ appears alone. But we have seen above that $K$ depends on the chosen system of units, and therefore an interpretation of the experiment in terms of a variation of constants is not meaningful (this is the case whether we consider a dimensionless $X$, like $\alpha$, or a dimensional one, like $m_e$).

Of course, this is not to say that there would not be an observable effect in the clock readings (indeed, in some cases there would be). What it does tell us, however, is that this non-local two-clock experiment is not sufficient to interpret the measurement in terms of a general variation of fundamental constants. Instead, such an experiment needs to be interpreted in terms of parameters of the underlying fundamental scalar field model.

As described in section \ref{sec:fund_const}, the one to one correspondence (\ref{eq:constants}) between coupling constants of the scalar field and fundamental constants is only valid in S.I. units, but can nonetheless be used as a useful tool to obtain results that are independent of the system of units. Using that correspondence and (\ref{eq:atom}) we can easily derive
\begin{equation}\label{dnu_nu_12_d}
\frac{\delta(\nu_{A1}/\nu_{A2})}{(\nu_{A1}/\nu_{A2})_0} = (2d_e+d_{m_e})\varphi(t,\bm x) .
\end{equation}

Although we obtained (\ref{dnu_nu_12_d}) by working in S.I. units, the result itself is independent of any system of units, as the dependencies on $d_i$ come from the fundamental Lagrangian (\ref{eq:Lint_lin}). Eq. (\ref{dnu_nu_12_d}) thus represents a meaningful experimental measurement. We make this more explicit by obtaining the same result working in atomic units in Appendix \ref{App:A}. 

Therefore, space-time separated clocks can provide meaningful  measurements (in the sense that they are independent of a conventional choice of units) of couplings between an underlying scalar field and SM fields. However, a meaningful interpretation in terms of space-time variation of constants is not possible. Such an interpretation will always depend on the system of units used. As a minimum, any such interpretation should explicitly specify the system of units it refers to. This is different from the more common case of co-located clock experiments, which are meaningful in both interpretations, measurement of some underlying scalar field, or space-time variation of fundamental constants.

Note that in (\ref{dnu_nu_12_d}) the constant $d_{m_e}$ appears alone. This is typical of space-time separated experiments, like \cite{GPSDM2017,Arvanitaki2018} and the one described here. In most other experiments\footnote{Exceptions are experiments using cavities with suspended mirrors, as proposed in \cite{StadnikLasInf2015,Geraci2018}.} analysed so far (see, e.g., \cite{Hees2018}) one only measures the combination $d_{m_e}-d_g$. In S.I. units, when interpreting the experiment in terms of space-time variation of constants, this corresponds to a measurement of the variation of $m_e$ (a dimensional constant) alone, rather than of the more usual dimensionless quantity $m_e/\Lambda_3$. But it is the variation of $m_e$ with respect to its value in a region where the scalar field is zero.
%But this is only valid in S.I. units i.e. in practice it corresponds to a measurement of the variation of $m_e$ with respect to the S.I. unit of mass which is implicitly assumed to not vary.

Thus one of the advantages of space time separated clock experiments is that they allow decorrelation of parameters ($d_{m_i}$ and $d_g$) that otherwise mostly appear as the combination $d_{m_e}-d_g$. 

\section{A time delayed clock comparison experiment} \label{sec:DAMNED}

We describe an experiment that compares the frequency of a clock (an ultra-stable optical cavity in this case) at time $t$ to its own frequency some time $t-T$ earlier, by ``storing" the output signal (photons) in a delay line. The advantages of such an experiment are two-fold: Firstly it allows searching for oscillations in the range 10-100~kHz, corresponding to DM masses of $4\times 10^{-11}$~eV to $4\times 10^{-10}$~eV, many orders of magnitude higher than usual clock based methods. Secondly, as described in the previous section, it is sensitive to a new combination of $d_{m_e}$ and $d_g$, and hence allows to break the degeneracy present in all co-located experiments, which always determine the same combination $d_{m_e}-d_g$.

\subsection{Experimental principle}
\begin{figure}[h!]
\includegraphics[width=0.5\textwidth]{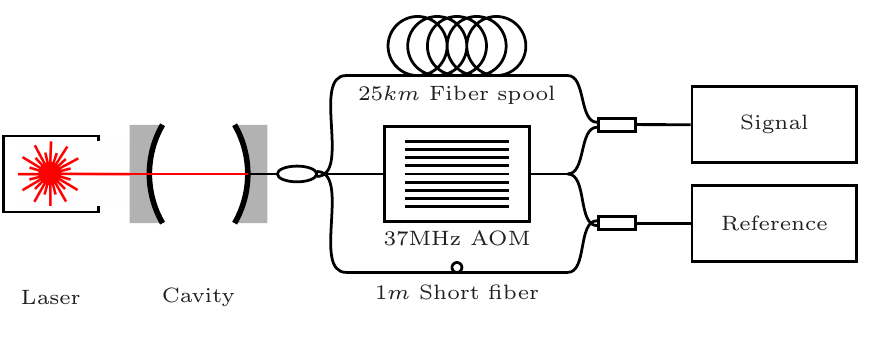}
\caption{Experimental setup. A $1542$nm laser source is locked to an ultra-stable cavity. The beam is then split between three arms and recombined to have access to the DM signal (long vs AOM arms) and the experimental reference (short vs AOM arms).}
\label{fig_Setup}
\end{figure}

Our experimental set-up, dubbed the DAMNED (DArk Matter from Non Equal Delays) experiment is a three-arm Mach-Zender interferometer as shown in figure \ref{fig_Setup}). A $1542$~nm laser source is stabilized on an ultra-stable  cavity \cite{Xie2017,Millo2009}, with a locking bandwidth of a few 100 kHz. The beam power is then unevenly distributed between the three arms. Most of the power is going through the long delay line that consists of a $25$~km fibre spool with a refractive index $n_0 \approx 1.5$. To perform a self-heterodyne detection, the laser frequency is shifted with the Acousto-optic modulator (AOM) located in the first arm (where $\nu_{AOM} = 37$~MHz). Finally, the last arm is a one meter fibre.

The beatnote between the AOM and the fibre spool arms provides the putative DM signal (see next section), with the reference beatnote between the AOM and the short fibre providing an indication of the experimental perturbations (noise and systematics) as the arm length is not sufficient to be affected by DM. Both beatnotes are acquired simultaneously using a two channel frequency counter (GuideTech668) at a sampling rate of $2.3$~MHz. The phase of the two beatnotes is computed from the counter readings and used for the DM analysis.

\subsection{Theoretical model} \label{sec:theo_mod}
In the theoretical framework discussed above, the cavity frequency $\omega$ as well as the fibre delay $T$ will oscillate at the Compton-de Broglie frequency $\omega_m$ corresponding to the DM mass. The scalar field at the location of the experiment is
\begin{equation} \label{eq:varphi(t)}
\varphi(t) = \varphi_0 \cos(\omega_m t)
\end{equation}
and the cavity length variation $\delta L(t) \equiv L(t) - L_0$ is
\begin{equation} \label{eq:L(t)}
\frac{\delta L(t)}{L_0} = - \epsilon_L \left((1+\alpha)\cos(\omega_mt) + \beta \sin(\omega_mt)\right)\, ,
\end{equation}
where $L_0$ is the unperturbed length, $\epsilon_L \equiv \varphi_0 (d_e+d_{m_e}) \ll 1$ is the fractional length change from the change of the Bohr radius (c.f. equ. (\ref{eq:cavity})) and where we neglect small ($\approx 10^{-4}d_i\varphi_0$) corrections coming from the $F_C$ term in (\ref{eq:cavity}) for our Si based cavity \cite{Pasteka2018}. The coefficients $\alpha,\beta$ are functions of the mechanical resonant frequencies $\omega_r$ of the cavity and become negligible when off resonance. At resonance $\beta \simeq Q$, the quality factor of our ULE cavity $Q \approx 6.1\times 10^4$ \cite{Millo2009,Numata2004,Zhang2013} and may therefore lead to significant enhancement of the signal. A detailed derivation of the coefficients $\alpha,\beta$ is provided in appendix \ref{App:B}.

The angular frequency variation $\delta \omega(t) \equiv \omega(t) - \omega_0$ of the light exiting the cavity is
\begin{equation} \label{eq:omega(t)}
\frac{\delta \omega(t)}{\omega_0} = \epsilon_L \left(\mathcal{E}_c(1+\alpha)\cos(\omega_mt) + \mathcal{E}_s \beta \sin(\omega_mt)\right)\, ,
\end{equation}
where $\omega_0$ is the unperturbed frequency. The coefficients $\mathcal{E}_c, \mathcal{E}_s$ are functions of the cavity finesse $\mathcal{F}$ and are related to the multiple passes of the light in the cavity. For our high finesse cavity ($\mathcal{F} \approx 800000$ \cite{Millo2009}) and frequencies of interest ($\omega_m \in [10,100]$~kHz) we have $\mathcal{E}_c,\mathcal{E}_s \simeq 1$. We provide a derivation of (\ref{eq:omega(t)}) with explicit expressions for $\mathcal{E}_c,\mathcal{E}_s$ based on \cite{Canuel2017,Vinet2010} in appendix \ref{App:C}.

The fibre delay is given by $T(t) = L_f(t)n(t)/c$, where $L_f(t)$ and $n(t)$ are the fibre length and refractive index respectively, which may both vary with the scalar field. We thus have
\begin{equation} \label{eq:dT_T_1}
\frac{\delta T(t)}{T_{0}} = \frac{\delta L_f(t)}{L_{f0}} + \frac{\delta n(t)}{n_{0}} \, .
\end{equation}
The length change will depend to leading order on the Bohr radius so $\delta L_f(t)/L_{f0} = -\epsilon_L\cos(\omega_m t)$, up to again $\approx 10^{-4}d_i\varphi_0$ corrections for our Si based fibre. The index change is a bit more involved, but can be related to the dispersion coefficient of the fibre and the frequency of the signal. Using the approach described in \cite{Braxmaier2001} we find, in S.I. units,
\begin{equation}\label{equ:n_Brax}
\frac{\delta n(t)}{n_0} = \frac{\omega_0}{n_0}\frac{\partial n}{\partial\omega}\left(\frac{\delta\omega(t)}{\omega_0}-2\frac{\delta\alpha(t)}{\alpha_0} - \frac{1}{2}\frac{\delta \mu(t)}{\mu_0} - \frac{\delta m_e(t)}{m_{e0}}\right)\, ,
\end{equation}
where we have defined $\mu \equiv m_e/m_N$ with $m_N$ the nucleon mass. For any nucleon (proton or neutron) one can decompose variations of $\mu$ to those of more fundamental quantities by $\frac{\delta \mu}{\mu_0} = \frac{\delta (m_e/\Lambda_3)}{(m_e/\Lambda_3)_0} - 0.048 \frac{\delta (m_q/\Lambda_3)}{(m_q/\Lambda_3)_0}$ (see, e.g., \cite{Flambaum2004}). Then (\ref{equ:n_Brax}) can be written
\begin{eqnarray}\label{equ:n_Brax2}
\frac{\delta n(t)}{n_0} &=& \frac{\omega_0}{n_0}\frac{\partial n}{\partial\omega}\left(\frac{\delta\omega(t)}{\omega_0} - \epsilon_n \cos(\omega_m t) \right) \, , 
\end{eqnarray}
where $\delta\omega(t)/\omega_0$ is given in (\ref{eq:omega(t)}), and $\epsilon_n \equiv \varphi_0 (2d_e + d_{m_e} + (d_{m_e}-d_g)/2 - 0.024(d_{m_q}-d_g))$. The pre-factor of (\ref{equ:n_Brax2}) depends on the refractive index and dispersion coefficient of the fibre, which can both be determined experimentally. For the telecom fibres that we use it is typically $\approx 10^{-2}$.

We can now write both, the cavity frequency and fibre delay, as a sum of two terms
\begin{eqnarray}
\frac{\delta \omega(t)}{\omega_{0}} &=& C_\omega \cos(\omega_m t) + S_\omega \sin(\omega_m t) \nonumber \\
\frac{\delta T(t)}{T_{0}} &=& C_T \cos(\omega_m t) + S_T \sin(\omega_m t)\, , 
\end{eqnarray}
where the small quantities $(C_T,S_T,C_\omega,S_\omega \ll 1)$ \footnote{For our range of frequencies $\varphi_0  \lesssim 2\times 10^{-20}$.} are obtained from (\ref{eq:omega(t)}), (\ref{eq:dT_T_1}) and (\ref{equ:n_Brax2}).

The propagation time of a signal arriving at the fibre output at time $t$ is then given to leading order by 

\begin{equation}
\begin{aligned}
T(t) = & \int_{t-T_0}^{t} \left[1 + C_T \cos(\omega_{m} t') + S_T \sin(\omega_{m} t')\right] \mathrm{d}t' \\
= &\, T_0 + 2 \frac{C_T}{\omega_{m}} \sin\left(\omega_{m} \frac{T_0}{2}\right) \cos\left(\omega_{m} t - \omega_{m} \frac{T_0}{2}\right) \\
& + 2 \frac{S_T}{\omega_{m}} \sin\left(\omega_{m} \frac{T_0}{2}\right) \sin\left(\omega_{m} t - \omega_{m} \frac{T_0}{2}\right) \, ,
\end{aligned}
\label{eq_T}
\end{equation}
and the phase difference between the delayed and non-delayed signals is
\begin{equation}
\begin{aligned}
&\Delta \Phi (t) = \int_{0}^{t} \omega(t')\, \mathrm{d} t' - \int_{0}^{t-T(t)} \omega(t')\, \mathrm{d} t' = \omega_0 T_0\\
&+ 2 \frac{\omega_{0} }{\omega_{m}} \sin\left(\omega_{m} \frac{T_0}{2}\right) \left[\left(C_T+C_\omega\right) \cos\left(\omega_{m} t - \omega_{m} \frac{T_0}{2}\right) \right.\\
& \hspace{3 cm} +\left.\left(S_T+S_\omega\right) \sin\left(\omega_{m} t - \omega_{m} \frac{T_0}{2}\right)\right] \, .
\end{aligned}
\label{eq_final}
\end{equation}
Note that for the reference arm $T_0 \approx 0$ and the last term representing the putative DM signal vanishes. Thus the ``Ref" signal (c.f. Fig. \ref{fig_Setup}) is a measure of all technical effects (noise and systematics) that are common to the ``Signal" and ``Reference" interferometers.

The DM signal we wish to detect is proportional to $\omega_0/\omega_m$ which indicates that, for a given phase measurement uncertainty, sensitivity improves with signal frequency $\omega_0$ favouring optical over e.g. microwave experiments. It is also proportional to $(C_T+C_\omega)$ and $(S_T+S_\omega)$, which are given by (\ref{eq:omega(t)}), (\ref{eq:dT_T_1}) and (\ref{equ:n_Brax2}):
\begin{equation} \label{eq:C-S}
\begin{aligned}
\left(C_T+C_\omega\right) &= \epsilon_L\left(\mathcal{E}_c(1+\alpha)\left(1+\frac{\omega_0}{n_0}\frac{\partial n}{\partial\omega}\right)-1\right) \\
&- \epsilon_n \frac{\omega_0}{n_0}\frac{\partial n}{\partial\omega}  \, , \\
\left(S_T+S_\omega\right) &= \epsilon_L\mathcal{E}_s\beta \left(1+ \frac{\omega_0}{n_0}\frac{\partial n}{\partial\omega}\right) \, .
\end{aligned}
\end{equation}
To link to the DM coupling constants we recall the definitions of $\epsilon_L$ and $\epsilon_n$
\begin{equation} \label{eq:epsilons}
\begin{aligned}
\epsilon_L &\equiv \varphi_0 (d_e+d_{m_e}) \\
\epsilon_n &\equiv \varphi_0 (2d_e + d_{m_e} + (d_{m_e}-d_g)/2 - 0.024(d_{m_q}-d_g)) \, .
\end{aligned}
\end{equation}

The coefficients $\mathcal{E}_c,\mathcal{E}_s$ are given explicitly in appendix \ref{App:C}. For our experiment we have $\mathcal{E}_c,\mathcal{E}_s \simeq 1$. The coefficients $\alpha, \beta$ are given explicitly in appendix \ref{App:B}. They can reach up to $\approx 6\times 10^4$ at resonance. For our 0.1~m ULE cavity the resonant frequencies are $\omega_r \approx 2 \pi i \, 19.6$~kHz where $i$ is an integer ($i\geq 1$), and are therefore well within our frequency region of interest ($[10,100]$~kHz. Finally, for our fibre $\frac{\omega_0}{n_0}\frac{\partial n}{\partial\omega} \approx 10^{-2}$.  

So in the presence of an oscillating scalar field we expect to see an oscillation of our measured phase difference in the signal port that is given by (\ref{eq_final}) and is related to the coupling constants $d_i$ by (\ref{eq:C-S}) and (\ref{eq:epsilons}). Finally, the amplitude of the field fluctuations $\varphi_0$ in (\ref{eq:epsilons}) is related to the DM density by (\ref{eq:phi_0}). Note that the signal in (\ref{eq_final}) goes to zero when the oscillation frequency is such that $\omega_m T_0/2 = j\pi$, with $j$ an integer. This limitation can be simply overcome by repeating the experiment with different lengths of fibre.

%Coming back to the discussion of section \ref{sec:clocks} we see in (\ref{eq:epsilons}) that now $d_{m_e}$ does not appear only in the combination $d_{m_e}-d_g$, typical of co-located experiments, but also by itself,  typical of space-time separated clock experiments, in our case a purely time-like separation. As a consequence our experiment, when combined with existing constraints \cite{Hees2018}, will allow decorrelating constraints on the two parameters.

\subsection{Some preliminary results} \label{sec:prelim_results}

Using the setup described above, the signal beatnote frequency $\nu_{S}$ and reference beatnote frequency $\nu_{R}$ are recorded simultaneously. In each case, we have a mean value centred around the AOM frequency ($\nu_{AOM}\approx 37$~MHz). We evaluate the relative frequency difference $y(t) = \left(\nu(t) - \nu_{AOM}\right)/\nu_{AOM}$ and then compute the one-sided power spectral density (PSD) of $y$, $S_{[y]}(f)$ from which we obtain the phase noise PSD $S_{[\Delta \Phi]}(f) = \left(\nu_{AOM}/f\right)^2 S_{[y]}(f)$.

\begin{figure}[h!]
\includegraphics[width=0.5\textwidth]{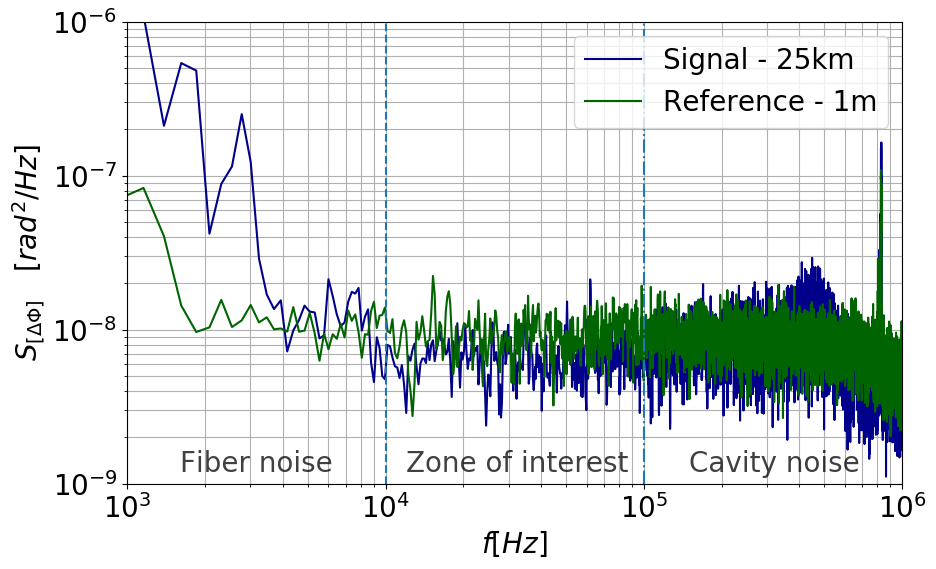}
\caption{Preliminary results for the PSD of phase fluctuations $S_{\Delta \Phi}(f)$ showing average values out of 12 runs. The setup is limited by the fibre noise below $10$~kHz, by the cavity phase noise above $100$~kHz and by the laser shot noise in our zone of interest.}
\label{fig_FR}
\end{figure}

Preliminary results can be seen in figure \ref{fig_FR}. They are obtained from 12 measurement runs of 4.32 ms, each comprising 10000 frequency measurements at 432 ns sampling. All measurements were taken within an interval of about 2 minutes, i.e., all under similar environmental conditions in the lab. 

The blue (resp. green) solid line is the average of the 12 PSDs from the signal (resp. reference) interferometer.

The PSDs delineate three frequency intervals :
\begin{itemize}
\item Below $10$~kHz, the PSD is limited by the acoustic and thermal noise of the long signal fibre. Note that this noise is absent in the reference fibre, as one would expect.
\item Above $100$~kHz, the PSD is limited by the short term stability of the laser and cavity combination (c.f. Fig. 4 of \cite{Xie2017}, laser B). One can clearly see the ``bump" of the PSD around 400~kHz coming from the cavity locking bandwidth. Again that is absent in the signal from the short fibre, as one would expect. 
\item Between $10$~kHz and $100$~kHz, the PSD is mainly limited by our measurement noise, as discussed below. This is our region of interest as the dominating noise is common to the signal and reference, and well understood.
\end{itemize}

The dominant noise in the region of interest is laser shot noise on the diodes. This is indicated by the white phase noise behaviour of the PSD in this region. To confirm that hypotheses, we have varied the incident power on the diode, and seen a linear dependence of the PSD level on the laser power.
%in the region 10-100~$\mu$W. The results of figure \ref{fig_FR} were obtained with a power of 25~$\mu$W, adjusted to be identical on the signal and reference outputs.

The maximum laser power (and min. shot noise) that we can use is limited in the present set-up by the stability of the cavity. Indeed, estimating the contribution from the cavity noise given in \cite{Xie2017} on our unequal arm interferometer we see that it is not far below the noise level shown in figure \ref{fig_FR}, at $\approx 1\times 10^{-9}$~rad$^2$/Hz. Indeed, at our highest optical powers we start seeing the characteristic ``bumps" arising from the laser noise multiplied by the transfer function of the unequal arm-length interferometer. This was particularly the case when the cavity was performing non-optimally, e.g., after a power outage and re-lock \footnote{One ``collateral" result of our experiment, as it turns out, is that we have built a rather sensitive ``real-time" analysis tool of the cavity performance at high frequency, which allows quick and unambiguous (in the sense that it does not rely on another reference cavity or external reference) identification of some cavity characteristics like locking bandwidth and performance.}.

Concerning the noise below $\sim 10$~kHz, it could probably be improved by placing the fibre spool in a controlled environment (temperature, acoustics), or even in vacuum if necessary. At present the fibre spool is in open air in the laboratory, which although not optimal, does not seem to contribute significantly in our main region of interest.

\subsection{Projected reach of the experiment}

We do not provide final results of our experiment yet, as we are still working on our long term data acquisition system and studying systematics. However, based on the observed noise level (see sect. \ref{sec:prelim_results}) we can provide rough estimates of the potential reach of our experiment.

We assume that the experiment is run with two fibre lengths (52.96 and 56.09 km), continuously for $T_{obs} = 10^6$~s ($\approx 11.6$~days) each. The two different fibre lengths are required to avoid the regions of zero sensitivity (see discussion at end of sect. \ref{sec:theo_mod}). We assume that the maximum amplitude of a potential harmonic signal at frequency $f$ that we can detect is given by
\begin{equation}
\Delta\Phi_{max}(f) = \sqrt{\frac{2S_{[\Delta \Phi]}(f)}{T_{obs}}} \, ,
\end{equation}
where $S_{[\Delta \Phi]}(f)$ is the (one-sided) PSD observed in our preliminary runs (see sect. \ref{sec:prelim_results} and fig. \ref{fig_FR}). Those maximum amplitudes can then be used directly in (\ref{eq_final}), (\ref{eq:C-S}) and (\ref{eq:epsilons}) to obtain the experimental reach in terms of the coupling parameters $d_i$ and as a function of the DM mass (or equivalently oscillation frequency $\omega_m$) for our relevant frequency range $\omega_m \in 2\pi[10,100]$~kHz.

\begin{figure}[h!]
\includegraphics[width=0.5\textwidth]{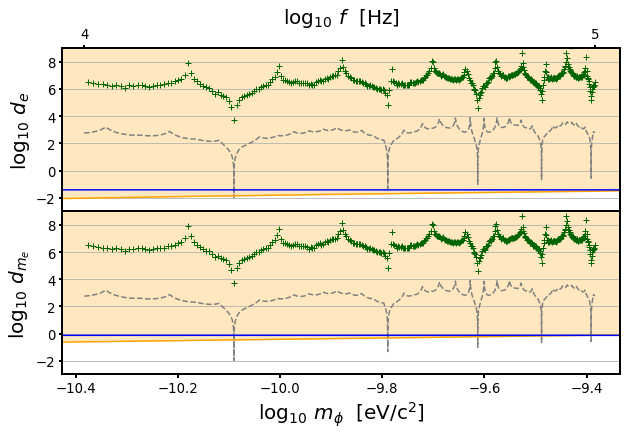}
\caption{Projected reach of our experiment on $d_e$ and $d_{m_e}$ (assuming all other $d_i = 0$ in turn) using existing data with $T_{obs} \approx 52$~ms (green crosses) and assuming a long run with $T_{obs} \approx 10^6$~s with the same noise (dashed grey line). The exiting data uses a single 25 km fibre spool. For the long run we intend to use two spools of 52.69~km and 56.09~km. One can see the advantages of a larger $T_{obs}$ that would improve the sensitivity, and of the combination of fibres that would smooth out the ``blind spots'' of our unequal-arm-length interferometer (effect of the $\sin(\omega_m T_0/2)$ term in (\ref{eq_final})). For comparison, best existing constraints extracted from \cite{Hees2018, Schlamminger2008, Smith1999} are also shown (blue and orange solid lines).}
\label{fig_de}
\end{figure}

For simplicity, we assume in turn that only one of the coupling parameters in (\ref{eq:epsilons}) is non-zero and thus give results on $d_e$ and $d_{m_e}$ independently. They are shown in figure \ref{fig_de}. We see that our experiment has the potential to improve on the only existing constraints, coming from tests of the weak equivalence principle (c.f. fig. 3 of \cite{Hees2018}, and \cite{Schlamminger2008,Smith1999}), by one to two orders of magnitude.\footnote{The constraints extracted from \cite{Schlamminger2008} in \cite{Hees2018} (orange lines in fig. \ref{fig_de}) should be used with caution for the DM masses here, which corresponds to a Yukawa range of $[0.5,5]$~km, meaning that a full modelling of the local mass distribution is required, well beyond the ``simple'' two-layer Earth model used in \cite{Hees2018}.} Additionally, we note that existing results can only constrain the combination $d_{m_e}-d_g$  (co-located experiments), whereas our experiment provides constraints on different combinations (as given in (\ref{eq:epsilons})) and thus should allow to completely disentangle $d_{m_e}$ from $d_g$ when combined with previous constraints.

Note that although based on real data, the results presented here do not take systematic effects into account nor is the data analysis optimized yet, so they should only be taken as our projected reach, not actual constraints.

%\footnote{Results on $d_{m_q}-d_g$ can be obtained from those of $d_{m_e}-d_g$ from a simple rescaling by a factor 1/0.048.}. At each frequency we take as our reach the lowest one obtained from the two quadratures of (\ref{eq_final}) and the two fibre lengths.

%\begin{figure}[h!]
%\includegraphics[width=0.5\textwidth]{DAMNED_dme}
%\caption{Projected reach of our experiment on $d_{m_e}$ (assuming all other $d_i = 0$). All other figure details identical to fig. \ref{fig_de}.}
%\label{fig_dme}
%\end{figure}

In summary, our experiment has the potential to detect scalar DM in spite of existing constraints from weak equivalence principle tests. This is particularly true at DM masses corresponding to the resonant frequencies of our cavity, but also elsewhere if some fine-tuning drives {\it e.g.} the combination $d_{m_e} - d_g$ to values that are about a factor $10^{3}$ smaller than $d_{m_e},d_g$ individually. Indeed, the possibility to decorrelate $d_{m_e}$ and $d_g$ is one of the main advantages of our space-time separated clock experiment.

\section{Conclusion and outlook}

We have presented a general theoretical analysis of space-time separated clock experiments in the context of a non-universally coupled massive scalar field that could be DM, and the related space-time variation of fundamental constants. Our main result is to demonstrate that although the interpretation of such experiments as a search for space-time variation of constants is not meaningful (in the sense that such an interpretation is dependent on a conventional choice of units), they still provide meaningful results when interpreted in a more fundamental scalar field theory. Consequently, they are capable of detecting DM if it comes in the form of such massive scalar fields. Additionally, such space-time separated clock experiments %contrary to co-located ones (except the proposal \cite{Geraci2018}), 
allow the breaking of the degeneracy between the couplings to fermion masses and gluons. More specifically, whilst most other experiments (except the proposal \cite{Geraci2018}) are sensitive to the linear combination of coupling constants $d_{m_i} - d_g$ (where $i = e,q$), space-time separated clock experiment are sensitive to $d_{m_i}$ alone.

Furthermore, we have described a novel experiment that is currently running at the Paris observatory, and is precisely such a space-time separated clock experiment. This experiment has two advantages: allowing decorrelation of $d_{m_e}$ and $d_g$ as described above, and allowing a much higher sampling rate than all other clock experiments, therefore exploring the high mass region of DM parameter space between $4 \times 10^{-11}$~eV and $4 \times 10^{-10}$~eV. We have presented first preliminary results from that experiment, and discussed its reach in a full DM search, showing that it has the potential to improve on existing constraints by one to two orders of magnitude.

In the upcoming months we will collect data in different configurations (different fibre lengths) and study in more detail the fundamental noise limits and systematic effects, as well as resonance effects of the cavity. We will also explore alternative configurations (e.g., different interferometer geometries, different fibres, different laser frequencies) that could give access to other parameter combinations and may allow an improvement of the performance.

We are also investigating other possibilities of DM detection. For example, in the mass range we are targeting, the coherence time of the DM oscillations is typically $\leq$~100~s so running the experiment for longer times will allow searching for the spectral profile of DM and its annual modulation \cite{DereviankoVULF2016,RobertsAsymm2018}. Furthermore, we are intending to use such an experiment for detection of transient DM events, as expected e.g. if DM forms topological defects \cite{DereviankoDM2014,GPSDM2018,Wcislo2016,Wcislo2018}. However, for a positive detection in that case, one requires several independent detectors for cross-correlation analyses, and we encourage other groups to set up similar experiments in view of a future network for such searches.\footnote{Alternatively, one can get a positive detection from long observation times with even a single device by looking for an annual modulation in DM-induced statistical signatures \cite{RobertsAsymm2018}.}

%We are also investigating the possibility to use such an experiment for detection of transient DM events, as expected e.g. if DM forms topological defects \cite{DereviankoDM2014,GPSDM2018,Wcislo2016,Wcislo2018}. However, for a positive detection in that case, one requires several independent detectors for cross-correlation analyses, and we encourage other groups to set up similar experiments in view of a future network for such searches.\footnote{Alternatively, one can get a positive detection from long observation time with even a single device by looking for an annual modulation in DM-induced statistical signatures \cite{RobertsAsymm2018}.}

\begin{acknowledgements}
We gratefully acknowledge help with the experiment by Etienne Cantin, and very useful discussions with Yevgeny Stadnik and Aur\'elien Hees. B.M.R. acknowledges financial support of Labex FIRST-TF. A.D. and C.D. acknowledge partial support of the U.S. National Science Foundation. This research was partially supported by the Australian Research Council Centre of Excellence for Engineered Quantum Systems (EQUS, CE170100009).
\end{acknowledgements}

\appendix
\section{}\label{App:A}

In this appendix, we derive explicitly the observable frequency variation due to the interaction (\ref{eq:L_au}) of atoms with the dark matter field $\varphi$.
As discussed in Sec.~\ref{sec:DM_gen}, the interactions with such scalar fields can be interpreted as an effective variation of fundamental constants.
This is a convenient parametrisation, since it allows us to leverage the extensive existing literature. However, such an interpretation is not universal (as it depends on the system of units employed), so care must be taken. Of course, the experimental results in terms of the scalar field couplings are independent of the choice of units.
%Of course, the physics and experimental results do not depend on the units. %...something..., and a sound physical interpretation

We derive the effective sensitivity coefficients that quantify the linear response of a given atomic transition to the $\varphi$-dependent perturbation, defined:
\begin{equation}
\frac{\delta\nu}{\nu} = \kappa_x \, d_x\varphi
\end{equation}
(with, e.g., $x=e$ or $m_e$, see Section~\ref{sec:DM_gen} above).
We denote these coefficients as $\kappa_x$ in order to distinguish them from the $K_X$ factors relevant to the case of general variation of fundamental constants.
Unlike the $K_X$ factors, the $\kappa_x$ coefficients are defined strictly via perturbation theory, and therefore do not depend on the units. They are thus well defined,  even for a single transition, allowing one to perform meaningful experiments with spatially (or temporally) separated clocks.

In order to obtain Eq. (\ref{dnu_nu_12_d}) in atomic units, we start from the quantum electrodynamics (QED) part of the Lagrangian density including the interaction terms from Eq.~(\ref{eq:L_au})

\begin{multline}\label{eq:LSMme-au}
\mathcal{L}_{{\rm SM}+{\rm int}}
	=  i \alpha^{-1} \bar\psi\gamma^\mu \partial_\mu\psi 
	- \alpha^{-2}\bar\psi\psi\left( 1 + d_{m_e}\varphi\right)  \\
	+ \alpha^{-1}\bar\psi\gamma^\mu A_\mu\psi 
	- \frac{\alpha^{-2}}{16\pi }F_{\mu\nu}F^{\mu\nu}\left( 1 - d_{e}\varphi\right).
\end{multline}
For now, we make the assumption that $\varphi$ varies slowly in space and time compared to the atomic size and time scales of the considered atomic transitions.
For the time-scales considered in this work (see Section \ref{sec:DAMNED}), this condition is easily satisfied.

In this regime, all derivatives of the $\varphi$ field vanish, and the resulting perturbative Hamiltonian can simply be derived in exact analogy with the regular ($\varphi=0$) case.
In the non-relativistic limit, this perturbation potential becomes
\begin{equation}\label{eq:dV}
\delta V =\varphi\left( d_{m_e} \alpha^{-2} + \frac{d_{m_e}}{2}\nabla^2 + d_{e} V(r)  \right),
\end{equation}
where $\nabla$ acts on electron coordinates, and $V(r)$ is the effective electrostatic potential. 
Note that we have written the potentials here in the single-particle form; for many-body systems there is also a summation over particles (inter-electron Coulomb interaction, as well as nuclear potential, included in $V(r)$).
Being constant, the first term in the parenthesis of Eq.~(\ref{eq:dV}) leads to no observable effects on atomic transition frequencies, and we will thus ignore it from here on
(this term may be interpreted as an effective addition to the electron inertia).

To calculate the resulting energy shifts, note that the non-relativistic perturbation (\ref{eq:dV}) can be re-written as
\begin{equation}
\delta V =\varphi\left(  d_{m_e} \left[V(r) - \hat H\right]  + d_{e} V(r) \right),
\end{equation}
where $H = \tfrac{1}{2}{\bm p}^2 + V$ is the non-relativistic Hamiltonian (in atomic units).
For a single-electron atom (H-like ion), we have $V(r) = -Z/r$, and noting $\langle{n}|Z r^{-1}|{n}\rangle =Z^2\, n^{-2} = -2E_n$ \cite{BetheBook}, we have:
\begin{equation}\label{eq:dE-H}
\delta E_n = \varphi(d_{m_e} + 2 d_e) E_n,
\end{equation}
corresponding to $\kappa_{m_e}=1$ and $\kappa_{e}=2$. Similar arguments can be made for more complex atomic systems.
For a (neutral) many-electron atom, $V(r)$ has the form $-Z/r$ for $r\ll a_0/Z$, and $-1/r$ for $r\gg a_0$, where $a_0$ is the Bohr radius.
From the virial theorem, we have
$
\langle V\rangle = -2\langle  \tfrac{1}{2}{\bm p}^2\rangle  = 2\langle H\rangle ,
$
which again leads to the same result: 
$
\langle \delta V\rangle = \varphi\left( 2d_e  + d_{m_e}  \right)\langle H\rangle.
$

We consider the comparison of the frequencies of two optical atomic clocks, $A1$ and $A2$.
%Are you thinking what I'm thinking A1? I think I am A2!
In analogy with Eq.~(\ref{eq:dnu_nu_X}), we express the variations in this comparison as a sum of terms that are linear in $(\varphi d_x)$.
If the two clocks are co-located (that is, the value of $\varphi$ can be taken to be the same at the location of both clocks), then we can express this as
\begin{equation}\label{eq:co-located}
\frac{\delta(\nu_{A1}/\nu_{A2})}{(\nu_{A1}/\nu_{A2})_0}
= \varphi \left([\kappa_e^{(A1)}-\kappa_e^{(A2)}]  d_e + [\kappa_{m_e}^{(A1)}-\kappa_{m_e}^{(A2)}]  d_{m_e}\right),
\end{equation}
%\begin{equation}
%\frac{\delta(\nu_{A1}/\nu_{A2})}{(\nu_{A1}/\nu_{A2})_0}
%= \left(\kappa_e^{(A1,A2)}  d_e + \kappa_{m_e}^{(A1,A2)} d_{m_e}\right),
%\end{equation}
where $\kappa_e$ and $\kappa_{m_e}$ are sensitivity coefficients that depend, in general, on the specifics of the atomic transitions considered.
However, if $\varphi$ takes a different value at each clock location
(as must typically be assumed for space-time separated atomic clock experiments), the expression differs.
Following section \ref{sec:ST-sep_clocks}, we consider here the extreme case, where $\varphi=\varphi_1$  at the location of the first clock, but $\varphi=0$ at the second.
This is particularly relevant for the case of transient effects \cite{DereviankoDM2014}.
Then, the observable frequency variation becomes
\begin{align}
\frac{\delta(\nu_{A1}/\nu_{A2})}{(\nu_{A1}/\nu_{A2})_0}
=\frac{\delta(\nu_{A1})}{(\nu_{A1})_0}
&=\left(\kappa_e d_e + \kappa_{m_e} d_{m_e}\right) \varphi_1 , \\
& = \left(2d_e +  d_{m_e}\right)\varphi_1 , 
\end{align}
where for (non-relativistic) optical transitions we have $\kappa_e=2$ and $\kappa_{m_e}=1$ from Eq.~(\ref{eq:dE-H}).
Of course, this is just the same result as Eq. (\ref{dnu_nu_12_d}), which was derived in S.I. units by analogy to variation of fundamental constants in Section \ref{sec:ST-sep_clocks}.

As demonstrated above, and widely considered in the literature, for co-located clocks, these interactions can be parameterised in terms of the variation of fundamental constants.
For space-time separated clocks, however, the interpretation in terms of a general variation of fundamental constants cannot be made unambiguously.
This can be seen in the fact that the $K_X$ factors [defined via Eq.(\ref{equ:def_K})] cannot be defined consistently independently from the units; see, e.g., \cite{KozlovBudker2018}.
However, when interpreting the results in terms of the perturbation by a specific external field such as in Eq.~(\ref{eq:L_au}), the ambiguity is removed, and the $\kappa_x$ factors are well-defined, even for a single transition.
This means exotic physics experiments performed by comparing two frequencies of spatially (or temporally) separated atomic clocks can indeed be unambiguously interpreted, unlike in the case of general variation of constants.

We note that we have so far neglected the relativistic and many-body effects.
However, as is clear from the cancellation in Eq.~(\ref{eq:co-located}), for co-located clock experiments that use the same type of clock transition (e.g., optical), it is in fact {\em only} these corrections that remain after this cancellation.
For heavy systems, the relativistic corrections are not so small, and can be important or even dominant for space-time separated clock experiments as well.
To that end, we note that the relativistic correction to the expectation value of the perturbation $\langle V \rangle \propto -\langle 1/r \rangle$ reproduces the 
same relativistic correction as in the case of variation of the fine-structure constant denoted $K_{\rm rel}$ in~\cite{DzuFlaWebPRL1999}.
%This is not surprising given the correspondence between these interactions and the variation of fundamental constants.

Crucially, we note that the difference between any two $\kappa$ factors is identical to the difference between the $K$ factors for the corresponding constants and transitions.
This is due to the exact one-to-one correspondence between the $\varphi$-induced perturbation~(\ref{eq:L_au}) and the case of general variation of fundamental constants in the case of co-located clock experiments.
Therefore, all $\kappa$ values can be recovered from the relevant $K$ values, provided a single $\kappa$ is known; e.g.,\ that for hydrogen as calculated here~(\ref{eq:dE-H}).
This is fortunate, since the $K$ values are typically not trivial to calculate for complex many-body systems, but are readily available for many atoms in the literature, see, e.g., Ref.~\cite{FlambaumCJP2009}.

For example, for optical transitions then, we have $\kappa_e = 2 + K_{\rm rel}$.
The `2' factor is the same for any (optical) transition;
in contrast, the $K_{\rm rel}$ factors depend strongly on atomic number $Z$ as well as on many-body electron effects  \cite{Dzuba1999,DzuFlaWebPRL1999,Angstmann2004}.
For other types of transitions the scaling is different; e.g., for hyperfine transitions relevant for microwave atomic clocks, the factor is $\kappa_e = 4+K_{\rm rel}^{\rm hf}$, as recently considered in Ref.~\cite{GPSDM2017}.

\section{}\label{App:B}

In this appendix we model the resonant cavity in the presence of a temporal oscillation of the fundamental constants. We generalise the approach of \cite{Arvanitaki2016} representing the cavity by a parametrically driven  damped harmonic oscillator
\begin{equation} \label{eq:harmonic_osc_L}
\ddot{L}(t) + \frac{\omega_r(t)}{Q(t)}\left(\dot{L}(t) -  \dot{L}_{eq}(t)\right) + \omega_r^2(t) \left(L(t) - L_{eq}(t)\right) = 0\, ,
\end{equation}
where $L(t)$ is the cavity length, $L_{eq}(t)$ is the equilibrium cavity length\footnote{It is deviations with respect to $L_{eq}$ that give rise to internal damping and restoring forces.}, $\omega_r(t)$ is the resonant frequency, and $Q(t)$ its quality factor. We neglect any external driving force (e.g. thermal noise) as our experiment is dominated by shot noise from the laser measurement. 

Note that the parameters ($L_{eq},\omega_r,Q$) of the harmonic oscillator described by (\ref{eq:harmonic_osc_L}) are themselves functions of time because of the temporal variation of the fundamental constants that they depend on. We will write them as $Q(t) = Q_0(1+\epsilon_Q \cos(\omega_mt))$, $\omega_r(t) = \omega_{r0}(1+\epsilon_\omega \cos(\omega_mt))$, $L_{eq}(t) =  L_{eq0}(1-\epsilon_L \cos(\omega_mt))$, with all $\epsilon \ll 1$. In particular the variation of the equilibrium length depends on the variation of the Bohr radius, with $\epsilon_L \equiv \varphi_0 (d_e+d_{m_e})$ as given in (\ref{eq:epsilons}).

We define the displacement $D(t) \equiv L(t)-L_{eq}(t)$ and rewrite (\ref{eq:harmonic_osc_L}) as
\begin{equation} \label{eq:harmonic_osc_D}
\ddot{D}(t) + \frac{\omega_r(t)}{Q(t)}\dot{D}(t) + \omega_r^2(t) D(t) = -\ddot{L}_{eq}(t)\, .
\end{equation} 

For our ULE cavity \cite{Millo2009} we have $L_{eq0} \approx 0.1$~m, $Q_0 \approx 6 \times 10^{4}$, and $\omega_{r0} \approx 2\pi i \, 19.6$~kHz where $i$ is an integer ($i\geq 1$). We have used those values to numerically solve (\ref{eq:harmonic_osc_D}) for our experimental configuration with the sinusoidal variation of the parameters. We found that in the steady state solution the contributions from $\epsilon_Q$ and $\epsilon_\omega$ are negligible with respect to those of $\epsilon_L$ (assuming that all $\epsilon$ are of similar order of magnitude). The system is thus well represented by a driven damped harmonic oscillator as already studied in \citep{Arvanitaki2016}
\begin{equation} \label{eq:harmonic_osc_D2}
\ddot{D}(t) + \frac{\omega_{r}}{Q_0}\dot{D}(t) + \omega_{r}^2 D(t) =-\epsilon_L L_0\, \omega_m^2 \cos(\omega_m t) \, .
\end{equation}
where $\omega_r, Q_0, L_0$ are now constants. The steady state solution of (\ref{eq:harmonic_osc_D2}) is
\begin{equation}
D(t) = -\epsilon_L L_0 \left(\alpha \cos(\omega_m t) + \beta \sin(\omega_m t) \right)
\end{equation}
with
\begin{equation}
\begin{array}{c c c c c}
\alpha &= \frac{Q_0^2 \omega_{m}^2 \left(\omega_r^2-\omega_{m}^2\right)}{Q_0^2 \left(\omega_r^2-\omega_{m}^2\right)^2+\omega_r^2 \omega_{m}^2} & & \beta  &= \frac{Q_0 \omega_r \omega_{m}^3}{Q_0^2 \left(\omega_r^2-\omega_{m}^2\right)^2+\omega_r^2 \omega_{m}^2} \, ,
\end{array}
\label{eq_resonance}
\end{equation}
and the total length variation of the cavity is 
\begin{equation}
\begin{aligned}
L(t) &= L_{eq}(t) + D(t) \\
&= L_0\left(1 - \epsilon_L(1 + \alpha)\cos(\omega_m t) - \epsilon_L\beta\sin(\omega_m t) \right)
\end{aligned}
\end{equation}
as in (\ref{eq:L(t)}).

At resonance ($\omega_m = \omega_r$) we have $\alpha = 0$ and $\beta = Q_0$. Below resonance ($\omega_m \ll \omega_r$) both $\beta,\alpha \simeq 0$. Above resonance ($\omega_m \gg \omega_r$) $\beta \simeq 0$ but $\alpha \simeq -1$, due to the presence of the $\omega_m^2$ factor in the ``driving force'' term (right hand side of (\ref{eq:harmonic_osc_D2})), meaning the the cavity can no longer follow the oscillations of the equilibrium length. 
%In any case, at the high frequency end in our range of interest ($[10,100]$~kHz) the signal is dominated by the successive resonances, multiples of 19.6~kHz, so the $\alpha \simeq -1$ term from lower lying resonances becomes negligible.  

\section{}\label{App:C}

The description of the resonant light field inside a Fabry-Perot cavity of oscillating length $L(t) = L_0 \cos(\omega_m t)$ has been treated extensively in the context of gravitational wave detectors like LIGO, Virgo, and more recently MIGA and described in detail in e.g. \cite{Vinet2010,Canuel2017}. Those analyses apply directly to our cavity and we only recall the main results, for details the reader is referred to the original papers.

We follow in particular the analysis in annex A of \cite{Canuel2017}, starting from equ. (35) of \cite{Canuel2017}, which gives the phase variation of the resonant light field exiting a cavity whose length is varying as $L(t) =  \zeta_c L_0 \cos(\omega_m t)$ (with $\zeta_c \ll 1$),
\begin{equation} \label{eq:phi(t)}
\begin{aligned}
\phi(t) \simeq \frac{2\zeta_c L_0 \omega_0 r^2}{c\, (r^4-2r^2\cos(2\nu)+1)} &\left((r^2-1)\cos(\nu)\cos(\omega_mt) \right.\\
 & \left. -(r^2+1)\sin(\nu)\sin(\omega_mt)\right)\, ,
\end{aligned}
\end{equation}
where $r$ is the reflection coefficient of the cavity mirrors and $\nu \equiv \omega_m L_0/c$. For our cavity with finesse $\mathcal{F} \approx 800000$ we have $1-r^2 \approx 4\times 10^{-6}$ $\left(\frac{r^2}{1-r^2} \simeq \mathcal{F}/\pi\right)$ and $\nu \approx [2,20] \times 10^{-5}$ for our frequency range of $[10,100]$~kHz, so we will neglect the first term in (\ref{eq:phi(t)}).

The fractional frequency variation ($\delta\omega(t)/\omega_0 = \dot{\phi}(t)/\omega_0$) is given by 
\begin{equation} \label{eq:cavity_cos}
\begin{aligned}
\frac{\delta\omega(t)}{\omega_0} = \frac{-2 \zeta_c \, \nu \, r^2(1+r^2)\sin(\nu)}{r^4-2r^2\cos(2\nu)+1} &\cos(\omega_mt)\, .
\end{aligned}
\end{equation}

The result for $L(t) =  \zeta_s L_0 \sin(\omega_m t)$ is simply obtained from (\ref{eq:cavity_cos}) by shifting $\omega_mt \rightarrow \omega_mt-\pi/2$ i.e. replacing $\cos(\omega_mt) \rightarrow \sin(\omega_mt)$ and $\zeta_c \rightarrow \zeta_s$.

Comparing to (\ref{eq:L(t)}) we identify $\zeta_c = -\epsilon_L(1+\alpha)$ and $\zeta_s = -\epsilon_L\beta$, and comparing to (\ref{eq:omega(t)}) we finally obtain
\begin{equation} \label{eq:Es-Ec}
\mathcal{E}_c=\mathcal{E}_s = \frac{2 \, \nu \, r^2(1+r^2)\sin(\nu)}{r^4-2r^2\cos(2\nu)+1} \, .
\end{equation}

Evaluating (\ref{eq:Es-Ec}) for our cavity and frequency range we have $\mathcal{E}_c,\mathcal{E}_s \in [0.991,0.99991]$ i.e. $\approx 1$.

%The fractional frequency variation ($\delta\omega(t)/\omega_0 = \dot{\phi}(t)/\omega_0$ is given by 
%\begin{equation} \label{eq:cavity_cos}
%\begin{aligned}
%\frac{\delta\omega(t)}{\omega_0} = \frac{2\zeta_c L_0 \omega_m r^2}{c\, (r^4-2r^2\cos(2\nu)+1)} &\left((1-r^2)\cos(\nu)\sin(\omega_mt) \right.\\
% & \left. -(1+r^2)\sin(\nu)\cos(\omega_mt)\right)\, .
%\end{aligned}
%\end{equation}
%
%The case for $L(t) =  \zeta_s L_0 \sin(\omega_m t)$ is simply obtained from (\ref{eq:cavity_cos}) by shifting $\omega_mt \rightarrow \omega_mt-\pi/2$ resulting in
%\begin{equation} \label{eq:cavity_sin}
%\begin{aligned}
%\frac{\delta\omega(t)}{\omega_0} = \frac{2\zeta_s L_0 \omega_m r^2}{c\, (r^4-2r^2\cos(2\nu)+1)} &\left(-(1-r^2)\cos(\nu)\cos(\omega_mt) \right.\\
% & \left. -(1+r^2)\sin(\nu)\sin(\omega_mt)\right)\, .
%\end{aligned}
%\end{equation}

\bibliography{DMclock-bibtexbib}

\begin{thebibliography}{43}%
\makeatletter
\providecommand \@ifxundefined [1]{%
 \@ifx{#1\undefined}
}%
\providecommand \@ifnum [1]{%
 \ifnum #1\expandafter \@firstoftwo
 \else \expandafter \@secondoftwo
 \fi
}%
\providecommand \@ifx [1]{%
 \ifx #1\expandafter \@firstoftwo
 \else \expandafter \@secondoftwo
 \fi
}%
\providecommand \natexlab [1]{#1}%
\providecommand \enquote  [1]{``#1''}%
\providecommand \bibnamefont  [1]{#1}%
\providecommand \bibfnamefont [1]{#1}%
\providecommand \citenamefont [1]{#1}%
\providecommand \href@noop [0]{\@secondoftwo}%
\providecommand \href [0]{\begingroup \@sanitize@url \@href}%
\providecommand \@href[1]{\@@startlink{#1}\@@href}%
\providecommand \@@href[1]{\endgroup#1\@@endlink}%
\providecommand \@sanitize@url [0]{\catcode `\\12\catcode `\$12\catcode
  `\&12\catcode `\#12\catcode `\^12\catcode `\_12\catcode `\%12\relax}%
\providecommand \@@startlink[1]{}%
\providecommand \@@endlink[0]{}%
\providecommand \url  [0]{\begingroup\@sanitize@url \@url }%
\providecommand \@url [1]{\endgroup\@href {#1}{\urlprefix }}%
\providecommand \urlprefix  [0]{URL }%
\providecommand \Eprint [0]{\href }%
\providecommand \doibase [0]{http://dx.doi.org/}%
\providecommand \selectlanguage [0]{\@gobble}%
\providecommand \bibinfo  [0]{\@secondoftwo}%
\providecommand \bibfield  [0]{\@secondoftwo}%
\providecommand \translation [1]{[#1]}%
\providecommand \BibitemOpen [0]{}%
\providecommand \bibitemStop [0]{}%
\providecommand \bibitemNoStop [0]{.\EOS\space}%
\providecommand \EOS [0]{\spacefactor3000\relax}%
\providecommand \BibitemShut  [1]{\csname bibitem#1\endcsname}%
\let\auto@bib@innerbib\@empty
%</preamble>
\bibitem [{\citenamefont {Bertone}\ and\ \citenamefont
  {Tait}(2018)}]{Bertone2018}%
  \BibitemOpen
  \bibfield  {author} {\bibinfo {author} {\bibfnamefont {G.}~\bibnamefont
  {Bertone}}\ and\ \bibinfo {author} {\bibfnamefont {T.~M.~P.}\ \bibnamefont
  {Tait}},\ }\href {\doibase 10.1038/s41586-018-0542-z} {\bibfield  {journal}
  {\bibinfo  {journal} {Nature}\ }\textbf {\bibinfo {volume} {562}},\ \bibinfo
  {pages} {51} (\bibinfo {year} {2018})},\ \Eprint
  {http://arxiv.org/abs/1810.01668} {arXiv:1810.01668} \BibitemShut {NoStop}%
\bibitem [{\citenamefont {Safronova}\ \emph {et~al.}(2018)\citenamefont
  {Safronova}, \citenamefont {Budker}, \citenamefont {DeMille}, \citenamefont
  {Kimball}, \citenamefont {Derevianko},\ and\ \citenamefont
  {Clark}}]{AtomicReview2017}%
  \BibitemOpen
  \bibfield  {author} {\bibinfo {author} {\bibfnamefont {M.~S.}\ \bibnamefont
  {Safronova}}, \bibinfo {author} {\bibfnamefont {D.}~\bibnamefont {Budker}},
  \bibinfo {author} {\bibfnamefont {D.}~\bibnamefont {DeMille}}, \bibinfo
  {author} {\bibfnamefont {D.~F.~J.}\ \bibnamefont {Kimball}}, \bibinfo
  {author} {\bibfnamefont {A.}~\bibnamefont {Derevianko}}, \ and\ \bibinfo
  {author} {\bibfnamefont {C.~W.}\ \bibnamefont {Clark}},\ }\href {\doibase
  10.1103/RevModPhys.90.025008} {\bibfield  {journal} {\bibinfo  {journal}
  {Rev. Mod. Phys.}\ }\textbf {\bibinfo {volume} {90}},\ \bibinfo {pages}
  {025008} (\bibinfo {year} {2018})}\BibitemShut {NoStop}%
\bibitem [{\citenamefont {Derevianko}\ and\ \citenamefont
  {Pospelov}(2014)}]{DereviankoDM2014}%
  \BibitemOpen
  \bibfield  {author} {\bibinfo {author} {\bibfnamefont {A.}~\bibnamefont
  {Derevianko}}\ and\ \bibinfo {author} {\bibfnamefont {M.}~\bibnamefont
  {Pospelov}},\ }\href {\doibase 10.1038/nphys3137} {\bibfield  {journal}
  {\bibinfo  {journal} {Nat. Phys.}\ }\textbf {\bibinfo {volume} {10}},\
  \bibinfo {pages} {933} (\bibinfo {year} {2014})}\BibitemShut {NoStop}%
\bibitem [{\citenamefont {Arvanitaki}\ \emph {et~al.}(2015)\citenamefont
  {Arvanitaki}, \citenamefont {Huang},\ and\ \citenamefont {{Van
  Tilburg}}}]{Arvanitaki2014}%
  \BibitemOpen
  \bibfield  {author} {\bibinfo {author} {\bibfnamefont {A.}~\bibnamefont
  {Arvanitaki}}, \bibinfo {author} {\bibfnamefont {J.}~\bibnamefont {Huang}}, \
  and\ \bibinfo {author} {\bibfnamefont {K.}~\bibnamefont {{Van Tilburg}}},\
  }\href {\doibase 10.1103/PhysRevD.91.015015} {\bibfield  {journal} {\bibinfo
  {journal} {Phys. Rev. D}\ }\textbf {\bibinfo {volume} {91}},\ \bibinfo
  {pages} {015015} (\bibinfo {year} {2015})}\BibitemShut {NoStop}%
\bibitem [{\citenamefont {Stadnik}\ and\ \citenamefont
  {Flambaum}(2015{\natexlab{a}})}]{StadnikDMalpha2015}%
  \BibitemOpen
  \bibfield  {author} {\bibinfo {author} {\bibfnamefont {Y.~V.}\ \bibnamefont
  {Stadnik}}\ and\ \bibinfo {author} {\bibfnamefont {V.~V.}\ \bibnamefont
  {Flambaum}},\ }\href {\doibase 10.1103/PhysRevLett.115.201301} {\bibfield
  {journal} {\bibinfo  {journal} {Phys. Rev. Lett.}\ }\textbf {\bibinfo
  {volume} {115}},\ \bibinfo {pages} {201301} (\bibinfo {year}
  {2015}{\natexlab{a}})}\BibitemShut {NoStop}%
\bibitem [{\citenamefont {{Van Tilburg}}\ \emph {et~al.}(2015)\citenamefont
  {{Van Tilburg}}, \citenamefont {Leefer}, \citenamefont {Bougas},\ and\
  \citenamefont {Budker}}]{Tilburg2015}%
  \BibitemOpen
  \bibfield  {author} {\bibinfo {author} {\bibfnamefont {K.}~\bibnamefont {{Van
  Tilburg}}}, \bibinfo {author} {\bibfnamefont {N.}~\bibnamefont {Leefer}},
  \bibinfo {author} {\bibfnamefont {L.}~\bibnamefont {Bougas}}, \ and\ \bibinfo
  {author} {\bibfnamefont {D.}~\bibnamefont {Budker}},\ }\href {\doibase
  10.1103/PhysRevLett.115.011802} {\bibfield  {journal} {\bibinfo  {journal}
  {Phys. Rev. Lett.}\ }\textbf {\bibinfo {volume} {115}},\ \bibinfo {pages}
  {011802} (\bibinfo {year} {2015})}\BibitemShut {NoStop}%
\bibitem [{\citenamefont {Hees}\ \emph {et~al.}(2016)\citenamefont {Hees},
  \citenamefont {Gu{\'{e}}na}, \citenamefont {Abgrall}, \citenamefont {Bize},\
  and\ \citenamefont {Wolf}}]{Hees2016}%
  \BibitemOpen
  \bibfield  {author} {\bibinfo {author} {\bibfnamefont {A.}~\bibnamefont
  {Hees}}, \bibinfo {author} {\bibfnamefont {J.}~\bibnamefont {Gu{\'{e}}na}},
  \bibinfo {author} {\bibfnamefont {M.}~\bibnamefont {Abgrall}}, \bibinfo
  {author} {\bibfnamefont {S.}~\bibnamefont {Bize}}, \ and\ \bibinfo {author}
  {\bibfnamefont {P.}~\bibnamefont {Wolf}},\ }\href {\doibase
  10.1103/PhysRevLett.117.061301} {\bibfield  {journal} {\bibinfo  {journal}
  {Phys. Rev. Lett.}\ }\textbf {\bibinfo {volume} {117}},\ \bibinfo {pages}
  {061301} (\bibinfo {year} {2016})}\BibitemShut {NoStop}%
\bibitem [{\citenamefont {Wcis{\l}o}\ \emph {et~al.}(2016)\citenamefont
  {Wcis{\l}o}, \citenamefont {Morzy{\'{n}}ski}, \citenamefont {Bober},
  \citenamefont {Cygan}, \citenamefont {Lisak}, \citenamefont {Ciury{\l}o},\
  and\ \citenamefont {Zawada}}]{Wcislo2016}%
  \BibitemOpen
  \bibfield  {author} {\bibinfo {author} {\bibfnamefont {P.}~\bibnamefont
  {Wcis{\l}o}}, \bibinfo {author} {\bibfnamefont {P.}~\bibnamefont
  {Morzy{\'{n}}ski}}, \bibinfo {author} {\bibfnamefont {M.}~\bibnamefont
  {Bober}}, \bibinfo {author} {\bibfnamefont {A.}~\bibnamefont {Cygan}},
  \bibinfo {author} {\bibfnamefont {D.}~\bibnamefont {Lisak}}, \bibinfo
  {author} {\bibfnamefont {R.}~\bibnamefont {Ciury{\l}o}}, \ and\ \bibinfo
  {author} {\bibfnamefont {M.}~\bibnamefont {Zawada}},\ }\href {\doibase
  10.1038/s41550-016-0009} {\bibfield  {journal} {\bibinfo  {journal} {Nat.
  Astron.}\ }\textbf {\bibinfo {volume} {1}},\ \bibinfo {pages} {0009}
  (\bibinfo {year} {2016})}\BibitemShut {NoStop}%
\bibitem [{\citenamefont {Roberts}\ \emph {et~al.}(2017)\citenamefont
  {Roberts}, \citenamefont {Blewitt}, \citenamefont {Dailey}, \citenamefont
  {Murphy}, \citenamefont {Pospelov}, \citenamefont {Rollings}, \citenamefont
  {Sherman}, \citenamefont {Williams},\ and\ \citenamefont
  {Derevianko}}]{GPSDM2017}%
  \BibitemOpen
  \bibfield  {author} {\bibinfo {author} {\bibfnamefont {B.~M.}\ \bibnamefont
  {Roberts}}, \bibinfo {author} {\bibfnamefont {G.}~\bibnamefont {Blewitt}},
  \bibinfo {author} {\bibfnamefont {C.}~\bibnamefont {Dailey}}, \bibinfo
  {author} {\bibfnamefont {M.}~\bibnamefont {Murphy}}, \bibinfo {author}
  {\bibfnamefont {M.}~\bibnamefont {Pospelov}}, \bibinfo {author}
  {\bibfnamefont {A.}~\bibnamefont {Rollings}}, \bibinfo {author}
  {\bibfnamefont {J.}~\bibnamefont {Sherman}}, \bibinfo {author} {\bibfnamefont
  {W.}~\bibnamefont {Williams}}, \ and\ \bibinfo {author} {\bibfnamefont
  {A.}~\bibnamefont {Derevianko}},\ }\href {\doibase
  10.1038/s41467-017-01440-4} {\bibfield  {journal} {\bibinfo  {journal} {Nat.
  Commun.}\ }\textbf {\bibinfo {volume} {8}},\ \bibinfo {pages} {1195}
  (\bibinfo {year} {2017})}\BibitemShut {NoStop}%
\bibitem [{\citenamefont {Hees}\ \emph {et~al.}(2018)\citenamefont {Hees},
  \citenamefont {Minazzoli}, \citenamefont {Savalle}, \citenamefont {Stadnik},\
  and\ \citenamefont {Wolf}}]{Hees2018}%
  \BibitemOpen
  \bibfield  {author} {\bibinfo {author} {\bibfnamefont {A.}~\bibnamefont
  {Hees}}, \bibinfo {author} {\bibfnamefont {O.}~\bibnamefont {Minazzoli}},
  \bibinfo {author} {\bibfnamefont {E.}~\bibnamefont {Savalle}}, \bibinfo
  {author} {\bibfnamefont {Y.~V.}\ \bibnamefont {Stadnik}}, \ and\ \bibinfo
  {author} {\bibfnamefont {P.}~\bibnamefont {Wolf}},\ }\href {\doibase
  10.1103/PhysRevD.98.064051} {\bibfield  {journal} {\bibinfo  {journal} {Phys.
  Rev. D}\ }\textbf {\bibinfo {volume} {98}},\ \bibinfo {pages} {064051}
  (\bibinfo {year} {2018})}\BibitemShut {NoStop}%
\bibitem [{\citenamefont {Wcis{\l}o}\ \emph {et~al.}(2018)\citenamefont
  {Wcis{\l}o}, \citenamefont {Ablewski}, \citenamefont {Beloy}, \citenamefont
  {Bilicki}, \citenamefont {Bober}, \citenamefont {Brown}, \citenamefont
  {Fasano}, \citenamefont {Ciury{\l}o}, \citenamefont {Hachisu}, \citenamefont
  {Ido}, \citenamefont {Lodewyck}, \citenamefont {Ludlow}, \citenamefont
  {McGrew}, \citenamefont {Morzy{\'n}ski}, \citenamefont {Nicolodi},
  \citenamefont {Schioppo}, \citenamefont {Sekido}, \citenamefont {Le~Targat},
  \citenamefont {Wolf}, \citenamefont {Zhang}, \citenamefont {Zjawin},\ and\
  \citenamefont {Zawada}}]{Wcislo2018}%
  \BibitemOpen
  \bibfield  {author} {\bibinfo {author} {\bibfnamefont {P.}~\bibnamefont
  {Wcis{\l}o}}, \bibinfo {author} {\bibfnamefont {P.}~\bibnamefont {Ablewski}},
  \bibinfo {author} {\bibfnamefont {K.}~\bibnamefont {Beloy}}, \bibinfo
  {author} {\bibfnamefont {S.}~\bibnamefont {Bilicki}}, \bibinfo {author}
  {\bibfnamefont {M.}~\bibnamefont {Bober}}, \bibinfo {author} {\bibfnamefont
  {R.}~\bibnamefont {Brown}}, \bibinfo {author} {\bibfnamefont
  {R.}~\bibnamefont {Fasano}}, \bibinfo {author} {\bibfnamefont
  {R.}~\bibnamefont {Ciury{\l}o}}, \bibinfo {author} {\bibfnamefont
  {H.}~\bibnamefont {Hachisu}}, \bibinfo {author} {\bibfnamefont
  {T.}~\bibnamefont {Ido}}, \bibinfo {author} {\bibfnamefont {J.}~\bibnamefont
  {Lodewyck}}, \bibinfo {author} {\bibfnamefont {A.}~\bibnamefont {Ludlow}},
  \bibinfo {author} {\bibfnamefont {W.}~\bibnamefont {McGrew}}, \bibinfo
  {author} {\bibfnamefont {P.}~\bibnamefont {Morzy{\'n}ski}}, \bibinfo {author}
  {\bibfnamefont {D.}~\bibnamefont {Nicolodi}}, \bibinfo {author}
  {\bibfnamefont {M.}~\bibnamefont {Schioppo}}, \bibinfo {author}
  {\bibfnamefont {M.}~\bibnamefont {Sekido}}, \bibinfo {author} {\bibfnamefont
  {R.}~\bibnamefont {Le~Targat}}, \bibinfo {author} {\bibfnamefont
  {P.}~\bibnamefont {Wolf}}, \bibinfo {author} {\bibfnamefont {X.}~\bibnamefont
  {Zhang}}, \bibinfo {author} {\bibfnamefont {B.}~\bibnamefont {Zjawin}}, \
  and\ \bibinfo {author} {\bibfnamefont {M.}~\bibnamefont {Zawada}},\
  }\href@noop {} {\bibfield  {journal} {\bibinfo  {journal} {Science Advances}\
  }\textbf {\bibinfo {volume} {4}},\ \bibinfo {pages} {eaau4869} (\bibinfo
  {year} {2018})}\BibitemShut {NoStop}%
\bibitem [{\citenamefont {Roberts}\ and\ \citenamefont
  {Derevianko}(2018)}]{RobertsAsymm2018}%
  \BibitemOpen
  \bibfield  {author} {\bibinfo {author} {\bibfnamefont {B.~M.}\ \bibnamefont
  {Roberts}}\ and\ \bibinfo {author} {\bibfnamefont {A.}~\bibnamefont
  {Derevianko}},\ }\href {http://arxiv.org/abs/1803.00617} {\  (\bibinfo {year}
  {2018})},\ \Eprint {http://arxiv.org/abs/1803.00617} {arXiv:1803.00617}
  \BibitemShut {NoStop}%
\bibitem [{\citenamefont {Roberts}\ \emph {et~al.}(2018)\citenamefont
  {Roberts}, \citenamefont {Blewitt}, \citenamefont {Dailey},\ and\
  \citenamefont {Derevianko}}]{GPSDM2018}%
  \BibitemOpen
  \bibfield  {author} {\bibinfo {author} {\bibfnamefont {B.~M.}\ \bibnamefont
  {Roberts}}, \bibinfo {author} {\bibfnamefont {G.}~\bibnamefont {Blewitt}},
  \bibinfo {author} {\bibfnamefont {C.}~\bibnamefont {Dailey}}, \ and\ \bibinfo
  {author} {\bibfnamefont {A.}~\bibnamefont {Derevianko}},\ }\href {\doibase
  10.1103/PhysRevD.97.083009} {\bibfield  {journal} {\bibinfo  {journal} {Phys.
  Rev. D}\ }\textbf {\bibinfo {volume} {97}},\ \bibinfo {pages} {083009}
  (\bibinfo {year} {2018})}\BibitemShut {NoStop}%
\bibitem [{\citenamefont {Alonso}\ \emph {et~al.}(2018)\citenamefont {Alonso},
  \citenamefont {Blas},\ and\ \citenamefont {Wolf}}]{Alonso2018}%
  \BibitemOpen
  \bibfield  {author} {\bibinfo {author} {\bibfnamefont {R.}~\bibnamefont
  {Alonso}}, \bibinfo {author} {\bibfnamefont {D.}~\bibnamefont {Blas}}, \ and\
  \bibinfo {author} {\bibfnamefont {P.}~\bibnamefont {Wolf}},\ }\href
  {http://arxiv.org/abs/1810.00889} {\  (\bibinfo {year} {2018})},\ \Eprint
  {http://arxiv.org/abs/1810.00889} {arXiv:1810.00889} \BibitemShut {NoStop}%
\bibitem [{\citenamefont {Wolf}\ \emph {et~al.}(2018)\citenamefont {Wolf},
  \citenamefont {Alonso},\ and\ \citenamefont {Blas}}]{Wolf2018}%
  \BibitemOpen
  \bibfield  {author} {\bibinfo {author} {\bibfnamefont {P.}~\bibnamefont
  {Wolf}}, \bibinfo {author} {\bibfnamefont {R.}~\bibnamefont {Alonso}}, \ and\
  \bibinfo {author} {\bibfnamefont {D.}~\bibnamefont {Blas}},\ }\href
  {http://arxiv.org/abs/1810.00889} {\  (\bibinfo {year} {2018})},\ \Eprint
  {http://arxiv.org/abs/1810.01632} {arXiv:1810.01632} \BibitemShut {NoStop}%
\bibitem [{\citenamefont {Arvanitaki}\ \emph {et~al.}(2018)\citenamefont
  {Arvanitaki}, \citenamefont {Graham}, \citenamefont {Hogan}, \citenamefont
  {Rajendran},\ and\ \citenamefont {Van~Tilburg}}]{Arvanitaki2018}%
  \BibitemOpen
  \bibfield  {author} {\bibinfo {author} {\bibfnamefont {A.}~\bibnamefont
  {Arvanitaki}}, \bibinfo {author} {\bibfnamefont {P.~W.}\ \bibnamefont
  {Graham}}, \bibinfo {author} {\bibfnamefont {J.~M.}\ \bibnamefont {Hogan}},
  \bibinfo {author} {\bibfnamefont {S.}~\bibnamefont {Rajendran}}, \ and\
  \bibinfo {author} {\bibfnamefont {K.}~\bibnamefont {Van~Tilburg}},\ }\href
  {\doibase 10.1103/PhysRevD.97.075020} {\bibfield  {journal} {\bibinfo
  {journal} {Phys. Rev. D}\ }\textbf {\bibinfo {volume} {97}},\ \bibinfo
  {pages} {075020} (\bibinfo {year} {2018})}\BibitemShut {NoStop}%
\bibitem [{\citenamefont {Uzan}(2003)}]{Uzan2003}%
  \BibitemOpen
  \bibfield  {author} {\bibinfo {author} {\bibfnamefont {J.-P.}\ \bibnamefont
  {Uzan}},\ }\href@noop {} {\bibfield  {journal} {\bibinfo  {journal} {Rev.
  Mod. Phys.}\ }\textbf {\bibinfo {volume} {75}},\ \bibinfo {pages} {403}
  (\bibinfo {year} {2003})}\BibitemShut {NoStop}%
\bibitem [{\citenamefont {Kozlov}\ and\ \citenamefont
  {Budker}(2018)}]{KozlovBudker2018}%
  \BibitemOpen
  \bibfield  {author} {\bibinfo {author} {\bibfnamefont {M.~G.}\ \bibnamefont
  {Kozlov}}\ and\ \bibinfo {author} {\bibfnamefont {D.}~\bibnamefont
  {Budker}},\ }\href {\doibase 10.1002/andp.201800254} {\bibfield  {journal}
  {\bibinfo  {journal} {Ann. Phys.}\ ,\ \bibinfo {pages} {1800254}} (\bibinfo
  {year} {2018})},\ \Eprint {http://arxiv.org/abs/1807.08337}
  {arXiv:1807.08337} \BibitemShut {NoStop}%
\bibitem [{\citenamefont {Damour}\ and\ \citenamefont
  {Donoghue}(2010)}]{Damour2010}%
  \BibitemOpen
  \bibfield  {author} {\bibinfo {author} {\bibfnamefont {T.}~\bibnamefont
  {Damour}}\ and\ \bibinfo {author} {\bibfnamefont {J.~F.}\ \bibnamefont
  {Donoghue}},\ }\href {\doibase 10.1103/PhysRevD.82.084033} {\bibfield
  {journal} {\bibinfo  {journal} {Phys. Rev. D}\ }\textbf {\bibinfo {volume}
  {82}},\ \bibinfo {pages} {084033} (\bibinfo {year} {2010})}\BibitemShut
  {NoStop}%
\bibitem [{\citenamefont {{Dzuba}}\ and\ \citenamefont
  {{Flambaum}}(2008)}]{dzuba:2008uq}%
  \BibitemOpen
  \bibfield  {author} {\bibinfo {author} {\bibfnamefont {V.~A.}\ \bibnamefont
  {{Dzuba}}}\ and\ \bibinfo {author} {\bibfnamefont {V.~V.}\ \bibnamefont
  {{Flambaum}}},\ }\href {\doibase 10.1103/PhysRevA.77.012515} {\bibfield
  {journal} {\bibinfo  {journal} {\pra}\ }\textbf {\bibinfo {volume} {77}},\
  \bibinfo {eid} {012515} (\bibinfo {year} {2008})},\ \Eprint
  {http://arxiv.org/abs/0712.3621} {arXiv:0712.3621 [physics.atom-ph]}
  \BibitemShut {NoStop}%
\bibitem [{\citenamefont {Flambaum}\ \emph {et~al.}(2004)\citenamefont
  {Flambaum}, \citenamefont {Leinweber}, \citenamefont {Thomas},\ and\
  \citenamefont {Young}}]{Flambaum2004}%
  \BibitemOpen
  \bibfield  {author} {\bibinfo {author} {\bibfnamefont {V.~V.}\ \bibnamefont
  {Flambaum}}, \bibinfo {author} {\bibfnamefont {D.~B.}\ \bibnamefont
  {Leinweber}}, \bibinfo {author} {\bibfnamefont {A.~W.}\ \bibnamefont
  {Thomas}}, \ and\ \bibinfo {author} {\bibfnamefont {R.~D.}\ \bibnamefont
  {Young}},\ }\href@noop {} {\bibfield  {journal} {\bibinfo  {journal} {Phys.
  Rev. D}\ }\textbf {\bibinfo {volume} {69}},\ \bibinfo {pages} {115006}
  (\bibinfo {year} {2004})}\BibitemShut {NoStop}%
\bibitem [{\citenamefont {Flambaum}\ and\ \citenamefont
  {Tedesco}(2006)}]{Flambaum2006}%
  \BibitemOpen
  \bibfield  {author} {\bibinfo {author} {\bibfnamefont {V.~V.}\ \bibnamefont
  {Flambaum}}\ and\ \bibinfo {author} {\bibfnamefont {A.~F.}\ \bibnamefont
  {Tedesco}},\ }\href@noop {} {\bibfield  {journal} {\bibinfo  {journal} {Phys.
  Rev. C}\ }\textbf {\bibinfo {volume} {73}},\ \bibinfo {pages} {055501}
  (\bibinfo {year} {2006})}\BibitemShut {NoStop}%
\bibitem [{\citenamefont {{McMillan}}(2011)}]{mcmillan:2011vn}%
  \BibitemOpen
  \bibfield  {author} {\bibinfo {author} {\bibfnamefont {P.~J.}\ \bibnamefont
  {{McMillan}}},\ }\href {\doibase 10.1111/j.1365-2966.2011.18564.x} {\bibfield
   {journal} {\bibinfo  {journal} {\mnras}\ }\textbf {\bibinfo {volume}
  {414}},\ \bibinfo {pages} {2446} (\bibinfo {year} {2011})},\ \Eprint
  {http://arxiv.org/abs/1102.4340} {arXiv:1102.4340 [astro-ph.GA]} \BibitemShut
  {NoStop}%
\bibitem [{\citenamefont {Bethe}\ and\ \citenamefont
  {Salpeter}(1977)}]{BetheBook}%
  \BibitemOpen
  \bibfield  {author} {\bibinfo {author} {\bibfnamefont {H.~A.}\ \bibnamefont
  {Bethe}}\ and\ \bibinfo {author} {\bibfnamefont {E.~E.}\ \bibnamefont
  {Salpeter}},\ }\href
  {http://wwwhep.physik.uni-freiburg.de/$\sim$fp/Versuche/FP2/FP2-10-Positronium/Anhang/H.A.Bethe,E.Salpeter.pdf}
  {\emph {\bibinfo {title} {{Quantum mechanics of one-and two-electron
  atoms}}}}\ (\bibinfo  {publisher} {Plenum Publishing Corporation},\ \bibinfo
  {address} {New York},\ \bibinfo {year} {1977})\BibitemShut {NoStop}%
\bibitem [{\citenamefont {Stadnik}\ and\ \citenamefont
  {Flambaum}(2015{\natexlab{b}})}]{Stadnik2015b}%
  \BibitemOpen
  \bibfield  {author} {\bibinfo {author} {\bibfnamefont {Y.~V.}\ \bibnamefont
  {Stadnik}}\ and\ \bibinfo {author} {\bibfnamefont {V.~V.}\ \bibnamefont
  {Flambaum}},\ }\href@noop {} {\bibfield  {journal} {\bibinfo  {journal}
  {Phys. Rev. Lett.}\ }\textbf {\bibinfo {volume} {114}},\ \bibinfo {pages}
  {161301} (\bibinfo {year} {2015}{\natexlab{b}})}\BibitemShut {NoStop}%
\bibitem [{\citenamefont {Pa{\v{s}}teka}\ \emph {et~al.}(2018)\citenamefont
  {Pa{\v{s}}teka}, \citenamefont {Hao}, \citenamefont {Borschevsky},
  \citenamefont {Flambaum},\ and\ \citenamefont {Schwerdtfeger}}]{Pasteka2018}%
  \BibitemOpen
  \bibfield  {author} {\bibinfo {author} {\bibfnamefont {L.~F.}\ \bibnamefont
  {Pa{\v{s}}teka}}, \bibinfo {author} {\bibfnamefont {Y.}~\bibnamefont {Hao}},
  \bibinfo {author} {\bibfnamefont {A.}~\bibnamefont {Borschevsky}}, \bibinfo
  {author} {\bibfnamefont {V.~V.}\ \bibnamefont {Flambaum}}, \ and\ \bibinfo
  {author} {\bibfnamefont {P.}~\bibnamefont {Schwerdtfeger}},\ }\href
  {http://arxiv.org/abs/1809.02863} {\  (\bibinfo {year} {2018})},\ \Eprint
  {http://arxiv.org/abs/1809.02863} {arXiv:1809.02863} \BibitemShut {NoStop}%
\bibitem [{\citenamefont {Stadnik}\ and\ \citenamefont
  {Flambaum}(2016)}]{StadnikLasInf2015}%
  \BibitemOpen
  \bibfield  {author} {\bibinfo {author} {\bibfnamefont {Y.~V.}\ \bibnamefont
  {Stadnik}}\ and\ \bibinfo {author} {\bibfnamefont {V.~V.}\ \bibnamefont
  {Flambaum}},\ }\href {\doibase 10.1103/PhysRevA.93.063630} {\bibfield
  {journal} {\bibinfo  {journal} {Phys. Rev. A}\ }\textbf {\bibinfo {volume}
  {93}},\ \bibinfo {pages} {063630} (\bibinfo {year} {2016})}\BibitemShut
  {NoStop}%
\bibitem [{\citenamefont {Geraci}\ \emph {et~al.}(2018)\citenamefont {Geraci},
  \citenamefont {Bradley}, \citenamefont {Gao}, \citenamefont {Weinstein},\
  and\ \citenamefont {Derevianko}}]{Geraci2018}%
  \BibitemOpen
  \bibfield  {author} {\bibinfo {author} {\bibfnamefont {A.~A.}\ \bibnamefont
  {Geraci}}, \bibinfo {author} {\bibfnamefont {C.}~\bibnamefont {Bradley}},
  \bibinfo {author} {\bibfnamefont {D.}~\bibnamefont {Gao}}, \bibinfo {author}
  {\bibfnamefont {J.}~\bibnamefont {Weinstein}}, \ and\ \bibinfo {author}
  {\bibfnamefont {A.}~\bibnamefont {Derevianko}},\ }\href
  {http://arxiv.org/abs/1808.00540} {\  (\bibinfo {year} {2018})},\ \Eprint
  {http://arxiv.org/abs/1808.00540} {arXiv:1808.00540} \BibitemShut {NoStop}%
\bibitem [{\citenamefont {Xie}\ \emph {et~al.}(2017)\citenamefont {Xie},
  \citenamefont {Bouchand}, \citenamefont {Nicolodi}, \citenamefont {Lours},
  \citenamefont {Alexandre},\ and\ \citenamefont {Coq}}]{Xie2017}%
  \BibitemOpen
  \bibfield  {author} {\bibinfo {author} {\bibfnamefont {X.}~\bibnamefont
  {Xie}}, \bibinfo {author} {\bibfnamefont {R.}~\bibnamefont {Bouchand}},
  \bibinfo {author} {\bibfnamefont {D.}~\bibnamefont {Nicolodi}}, \bibinfo
  {author} {\bibfnamefont {M.}~\bibnamefont {Lours}}, \bibinfo {author}
  {\bibfnamefont {C.}~\bibnamefont {Alexandre}}, \ and\ \bibinfo {author}
  {\bibfnamefont {Y.~L.}\ \bibnamefont {Coq}},\ }\href {\doibase
  10.1364/OL.42.001217} {\bibfield  {journal} {\bibinfo  {journal} {Opt.
  Lett.}\ }\textbf {\bibinfo {volume} {42}},\ \bibinfo {pages} {1217} (\bibinfo
  {year} {2017})}\BibitemShut {NoStop}%
\bibitem [{\citenamefont {Millo}\ \emph {et~al.}(2009)\citenamefont {Millo},
  \citenamefont {Magalh\~aes}, \citenamefont {Mandache}, \citenamefont
  {Le~Coq}, \citenamefont {English}, \citenamefont {Westergaard}, \citenamefont
  {Lodewyck}, \citenamefont {Bize}, \citenamefont {Lemonde},\ and\
  \citenamefont {Santarelli}}]{Millo2009}%
  \BibitemOpen
  \bibfield  {author} {\bibinfo {author} {\bibfnamefont {J.}~\bibnamefont
  {Millo}}, \bibinfo {author} {\bibfnamefont {D.~V.}\ \bibnamefont
  {Magalh\~aes}}, \bibinfo {author} {\bibfnamefont {C.}~\bibnamefont
  {Mandache}}, \bibinfo {author} {\bibfnamefont {Y.}~\bibnamefont {Le~Coq}},
  \bibinfo {author} {\bibfnamefont {E.~M.~L.}\ \bibnamefont {English}},
  \bibinfo {author} {\bibfnamefont {P.~G.}\ \bibnamefont {Westergaard}},
  \bibinfo {author} {\bibfnamefont {J.}~\bibnamefont {Lodewyck}}, \bibinfo
  {author} {\bibfnamefont {S.}~\bibnamefont {Bize}}, \bibinfo {author}
  {\bibfnamefont {P.}~\bibnamefont {Lemonde}}, \ and\ \bibinfo {author}
  {\bibfnamefont {G.}~\bibnamefont {Santarelli}},\ }\href {\doibase
  10.1103/PhysRevA.79.053829} {\bibfield  {journal} {\bibinfo  {journal} {Phys.
  Rev. A}\ }\textbf {\bibinfo {volume} {79}},\ \bibinfo {pages} {053829}
  (\bibinfo {year} {2009})}\BibitemShut {NoStop}%
\bibitem [{\citenamefont {Numata}\ \emph {et~al.}(2004)\citenamefont {Numata},
  \citenamefont {Kemery},\ and\ \citenamefont {Camp}}]{Numata2004}%
  \BibitemOpen
  \bibfield  {author} {\bibinfo {author} {\bibfnamefont {K.}~\bibnamefont
  {Numata}}, \bibinfo {author} {\bibfnamefont {A.}~\bibnamefont {Kemery}}, \
  and\ \bibinfo {author} {\bibfnamefont {J.}~\bibnamefont {Camp}},\ }\href
  {\doibase 10.1103/PhysRevLett.93.250602} {\bibfield  {journal} {\bibinfo
  {journal} {Phys. Rev. Lett.}\ }\textbf {\bibinfo {volume} {93}},\ \bibinfo
  {pages} {250602} (\bibinfo {year} {2004})}\BibitemShut {NoStop}%
\bibitem [{\citenamefont {Zhang}\ \emph {et~al.}(2013)\citenamefont {Zhang},
  \citenamefont {Luo}, \citenamefont {Ouyang}, \citenamefont {Deng},
  \citenamefont {Lu},\ and\ \citenamefont {Luo}}]{Zhang2013}%
  \BibitemOpen
  \bibfield  {author} {\bibinfo {author} {\bibfnamefont {J.}~\bibnamefont
  {Zhang}}, \bibinfo {author} {\bibfnamefont {Y.}~\bibnamefont {Luo}}, \bibinfo
  {author} {\bibfnamefont {B.}~\bibnamefont {Ouyang}}, \bibinfo {author}
  {\bibfnamefont {K.}~\bibnamefont {Deng}}, \bibinfo {author} {\bibfnamefont
  {Z.}~\bibnamefont {Lu}}, \ and\ \bibinfo {author} {\bibfnamefont
  {J.}~\bibnamefont {Luo}},\ }\href {\doibase 10.1140/epjd/e2013-30458-2}
  {\bibfield  {journal} {\bibinfo  {journal} {The European Physical Journal D}\
  }\textbf {\bibinfo {volume} {67}},\ \bibinfo {pages} {46} (\bibinfo {year}
  {2013})}\BibitemShut {NoStop}%
\bibitem [{\citenamefont {{Canuel}}\ \emph {et~al.}(2018)\citenamefont
  {{Canuel}}, \citenamefont {{Bertoldi}}, \citenamefont {{Amand}},
  \citenamefont {{Borgo di Pozzo}}, \citenamefont {{Fang}}, \citenamefont
  {{Geiger}}, \citenamefont {{Gillot}}, \citenamefont {{Henry}}, \citenamefont
  {{Hinderer}}, \citenamefont {{Holleville}}, \citenamefont {{Lef{\`e}vre}},
  \citenamefont {{Merzougui}}, \citenamefont {{Mielec}}, \citenamefont
  {{Monfret}}, \citenamefont {{Pelisson}}, \citenamefont {{Prevedelli}},
  \citenamefont {{Reynaud}}, \citenamefont {{Riou}}, \citenamefont
  {{Rogister}}, \citenamefont {{Rosat}}, \citenamefont {{Cormier}},
  \citenamefont {{Landragin}}, \citenamefont {{Chaibi}}, \citenamefont
  {{Gaffet}},\ and\ \citenamefont {{Bouyer}}}]{Canuel2017}%
  \BibitemOpen
  \bibfield  {author} {\bibinfo {author} {\bibfnamefont {B.}~\bibnamefont
  {{Canuel}}}, \bibinfo {author} {\bibfnamefont {A.}~\bibnamefont
  {{Bertoldi}}}, \bibinfo {author} {\bibfnamefont {L.}~\bibnamefont {{Amand}}},
  \bibinfo {author} {\bibfnamefont {E.}~\bibnamefont {{Borgo di Pozzo}}},
  \bibinfo {author} {\bibfnamefont {B.}~\bibnamefont {{Fang}}}, \bibinfo
  {author} {\bibfnamefont {R.}~\bibnamefont {{Geiger}}}, \bibinfo {author}
  {\bibfnamefont {J.}~\bibnamefont {{Gillot}}}, \bibinfo {author}
  {\bibfnamefont {S.}~\bibnamefont {{Henry}}}, \bibinfo {author} {\bibfnamefont
  {J.}~\bibnamefont {{Hinderer}}}, \bibinfo {author} {\bibfnamefont
  {D.}~\bibnamefont {{Holleville}}}, \bibinfo {author} {\bibfnamefont
  {G.}~\bibnamefont {{Lef{\`e}vre}}}, \bibinfo {author} {\bibfnamefont
  {M.}~\bibnamefont {{Merzougui}}}, \bibinfo {author} {\bibfnamefont
  {N.}~\bibnamefont {{Mielec}}}, \bibinfo {author} {\bibfnamefont
  {T.}~\bibnamefont {{Monfret}}}, \bibinfo {author} {\bibfnamefont
  {S.}~\bibnamefont {{Pelisson}}}, \bibinfo {author} {\bibfnamefont
  {M.}~\bibnamefont {{Prevedelli}}}, \bibinfo {author} {\bibfnamefont
  {S.}~\bibnamefont {{Reynaud}}}, \bibinfo {author} {\bibfnamefont
  {I.}~\bibnamefont {{Riou}}}, \bibinfo {author} {\bibfnamefont
  {Y.}~\bibnamefont {{Rogister}}}, \bibinfo {author} {\bibfnamefont
  {S.}~\bibnamefont {{Rosat}}}, \bibinfo {author} {\bibfnamefont
  {E.}~\bibnamefont {{Cormier}}}, \bibinfo {author} {\bibfnamefont
  {A.}~\bibnamefont {{Landragin}}}, \bibinfo {author} {\bibfnamefont
  {W.}~\bibnamefont {{Chaibi}}}, \bibinfo {author} {\bibfnamefont
  {S.}~\bibnamefont {{Gaffet}}}, \ and\ \bibinfo {author} {\bibfnamefont
  {P.}~\bibnamefont {{Bouyer}}},\ }\href {\doibase 10.1038/s41598-018-32165-z}
  {\bibfield  {journal} {\bibinfo  {journal} {Scientific Reports}\ }\textbf
  {\bibinfo {volume} {8}},\ \bibinfo {pages} {14064} (\bibinfo {year}
  {2018})},\ \Eprint {http://arxiv.org/abs/1703.02490} {arXiv:1703.02490
  [physics.atom-ph]} \BibitemShut {NoStop}%
\bibitem [{\citenamefont {Virgo-collaboration}(2010)}]{Vinet2010}%
  \BibitemOpen
  \bibfield  {author} {\bibinfo {author} {\bibnamefont {Virgo-collaboration}},\
  }\href
  {https://www.ego-gw.it/public/events/vesf/2010/Presentations/Interferometer_Materials-Vinet.pdf}
  {\emph {\bibinfo {title} {The VIRGO Physics Book, Vol. II, OPTICS and related
  TOPICS}}}\ (\bibinfo  {publisher} {The Virgo collaboration},\ \bibinfo {year}
  {2010})\BibitemShut {NoStop}%
\bibitem [{\citenamefont {Braxmaier}\ \emph {et~al.}(2001)\citenamefont
  {Braxmaier}, \citenamefont {Pradl}, \citenamefont {M\"uller}, \citenamefont
  {Peters}, \citenamefont {Mlynek}, \citenamefont {Loriette},\ and\
  \citenamefont {Schiller}}]{Braxmaier2001}%
  \BibitemOpen
  \bibfield  {author} {\bibinfo {author} {\bibfnamefont {C.}~\bibnamefont
  {Braxmaier}}, \bibinfo {author} {\bibfnamefont {O.}~\bibnamefont {Pradl}},
  \bibinfo {author} {\bibfnamefont {H.}~\bibnamefont {M\"uller}}, \bibinfo
  {author} {\bibfnamefont {A.}~\bibnamefont {Peters}}, \bibinfo {author}
  {\bibfnamefont {J.}~\bibnamefont {Mlynek}}, \bibinfo {author} {\bibfnamefont
  {V.}~\bibnamefont {Loriette}}, \ and\ \bibinfo {author} {\bibfnamefont
  {S.}~\bibnamefont {Schiller}},\ }\href {\doibase 10.1103/PhysRevD.64.042001}
  {\bibfield  {journal} {\bibinfo  {journal} {Phys. Rev. D}\ }\textbf {\bibinfo
  {volume} {64}},\ \bibinfo {pages} {042001} (\bibinfo {year}
  {2001})}\BibitemShut {NoStop}%
\bibitem [{\citenamefont {Schlamminger}\ \emph {et~al.}(2008)\citenamefont
  {Schlamminger}, \citenamefont {Choi}, \citenamefont {Wagner}, \citenamefont
  {Gundlach},\ and\ \citenamefont {Adelberger}}]{Schlamminger2008}%
  \BibitemOpen
  \bibfield  {author} {\bibinfo {author} {\bibfnamefont {S.}~\bibnamefont
  {Schlamminger}}, \bibinfo {author} {\bibfnamefont {K.-Y.}\ \bibnamefont
  {Choi}}, \bibinfo {author} {\bibfnamefont {T.~A.}\ \bibnamefont {Wagner}},
  \bibinfo {author} {\bibfnamefont {J.}~\bibnamefont {Gundlach}}, \ and\
  \bibinfo {author} {\bibfnamefont {E.~G.}\ \bibnamefont {Adelberger}},\
  }\href@noop {} {\bibfield  {journal} {\bibinfo  {journal} {Phys. Rev. Lett}\
  }\textbf {\bibinfo {volume} {100}},\ \bibinfo {pages} {041101} (\bibinfo
  {year} {2008})}\BibitemShut {NoStop}%
\bibitem [{\citenamefont {Smith}\ \emph {et~al.}(1999)\citenamefont {Smith},
  \citenamefont {Hoyle}, \citenamefont {Gundlach}, \citenamefont {Adelberger},
  \citenamefont {Heckel},\ and\ \citenamefont {Swanson}}]{Smith1999}%
  \BibitemOpen
  \bibfield  {author} {\bibinfo {author} {\bibfnamefont {G.~L.}\ \bibnamefont
  {Smith}}, \bibinfo {author} {\bibfnamefont {C.~D.}\ \bibnamefont {Hoyle}},
  \bibinfo {author} {\bibfnamefont {J.~H.}\ \bibnamefont {Gundlach}}, \bibinfo
  {author} {\bibfnamefont {E.~G.}\ \bibnamefont {Adelberger}}, \bibinfo
  {author} {\bibfnamefont {B.~R.}\ \bibnamefont {Heckel}}, \ and\ \bibinfo
  {author} {\bibfnamefont {H.~E.}\ \bibnamefont {Swanson}},\ }\href {\doibase
  10.1103/PhysRevD.61.022001} {\bibfield  {journal} {\bibinfo  {journal} {Phys.
  Rev. D}\ }\textbf {\bibinfo {volume} {61}},\ \bibinfo {pages} {022001}
  (\bibinfo {year} {1999})}\BibitemShut {NoStop}%
\bibitem [{\citenamefont {Derevianko}(2018)}]{DereviankoVULF2016}%
  \BibitemOpen
  \bibfield  {author} {\bibinfo {author} {\bibfnamefont {A.}~\bibnamefont
  {Derevianko}},\ }\href {\doibase 10.1103/PhysRevA.97.042506} {\bibfield
  {journal} {\bibinfo  {journal} {Phys. Rev. A}\ }\textbf {\bibinfo {volume}
  {97}},\ \bibinfo {pages} {042506} (\bibinfo {year} {2018})}\BibitemShut
  {NoStop}%
\bibitem [{\citenamefont {Dzuba}\ \emph
  {et~al.}(1999{\natexlab{a}})\citenamefont {Dzuba}, \citenamefont {Flambaum},\
  and\ \citenamefont {Webb}}]{DzuFlaWebPRL1999}%
  \BibitemOpen
  \bibfield  {author} {\bibinfo {author} {\bibfnamefont {V.~A.}\ \bibnamefont
  {Dzuba}}, \bibinfo {author} {\bibfnamefont {V.~V.}\ \bibnamefont {Flambaum}},
  \ and\ \bibinfo {author} {\bibfnamefont {J.~K.}\ \bibnamefont {Webb}},\
  }\href {\doibase 10.1103/PhysRevLett.82.888} {\bibfield  {journal} {\bibinfo
  {journal} {Phys. Rev. Lett.}\ }\textbf {\bibinfo {volume} {82}},\ \bibinfo
  {pages} {888} (\bibinfo {year} {1999}{\natexlab{a}})}\BibitemShut {NoStop}%
\bibitem [{\citenamefont {Flambaum}\ and\ \citenamefont
  {Dzuba}(2009)}]{FlambaumCJP2009}%
  \BibitemOpen
  \bibfield  {author} {\bibinfo {author} {\bibfnamefont {V.~V.}\ \bibnamefont
  {Flambaum}}\ and\ \bibinfo {author} {\bibfnamefont {V.~A.}\ \bibnamefont
  {Dzuba}},\ }\href {\doibase 10.1139/p08-072} {\bibfield  {journal} {\bibinfo
  {journal} {Can. J. Phys.}\ }\textbf {\bibinfo {volume} {87}},\ \bibinfo
  {pages} {25} (\bibinfo {year} {2009})}\BibitemShut {NoStop}%
\bibitem [{\citenamefont {Dzuba}\ \emph
  {et~al.}(1999{\natexlab{b}})\citenamefont {Dzuba}, \citenamefont {Flambaum},\
  and\ \citenamefont {Webb}}]{Dzuba1999}%
  \BibitemOpen
  \bibfield  {author} {\bibinfo {author} {\bibfnamefont {V.~A.}\ \bibnamefont
  {Dzuba}}, \bibinfo {author} {\bibfnamefont {V.~V.}\ \bibnamefont {Flambaum}},
  \ and\ \bibinfo {author} {\bibfnamefont {J.~K.}\ \bibnamefont {Webb}},\
  }\href {\doibase 10.1103/PhysRevA.59.230} {\bibfield  {journal} {\bibinfo
  {journal} {Phys. Rev. A}\ }\textbf {\bibinfo {volume} {59}},\ \bibinfo
  {pages} {230} (\bibinfo {year} {1999}{\natexlab{b}})}\BibitemShut {NoStop}%
\bibitem [{\citenamefont {Angstmann}\ \emph {et~al.}(2004)\citenamefont
  {Angstmann}, \citenamefont {Dzuba},\ and\ \citenamefont
  {Flambaum}}]{Angstmann2004}%
  \BibitemOpen
  \bibfield  {author} {\bibinfo {author} {\bibfnamefont {E.~J.}\ \bibnamefont
  {Angstmann}}, \bibinfo {author} {\bibfnamefont {V.~A.}\ \bibnamefont
  {Dzuba}}, \ and\ \bibinfo {author} {\bibfnamefont {V.~V.}\ \bibnamefont
  {Flambaum}},\ }\href {\doibase 10.1103/PhysRevA.70.014102} {\bibfield
  {journal} {\bibinfo  {journal} {Phys. Rev. A}\ }\textbf {\bibinfo {volume}
  {70}},\ \bibinfo {pages} {014102} (\bibinfo {year} {2004})}\BibitemShut
  {NoStop}%
\bibitem [{\citenamefont {Arvanitaki}\ \emph {et~al.}(2016)\citenamefont
  {Arvanitaki}, \citenamefont {Dimopoulos},\ and\ \citenamefont {{Van
  Tilburg}}}]{Arvanitaki2016}%
  \BibitemOpen
  \bibfield  {author} {\bibinfo {author} {\bibfnamefont {A.}~\bibnamefont
  {Arvanitaki}}, \bibinfo {author} {\bibfnamefont {S.}~\bibnamefont
  {Dimopoulos}}, \ and\ \bibinfo {author} {\bibfnamefont {K.}~\bibnamefont
  {{Van Tilburg}}},\ }\href {\doibase 10.1103/PhysRevLett.116.031102}
  {\bibfield  {journal} {\bibinfo  {journal} {Phys. Rev. Lett.}\ }\textbf
  {\bibinfo {volume} {116}},\ \bibinfo {pages} {031102} (\bibinfo {year}
  {2016})},\ \Eprint {http://arxiv.org/abs/1508.01798} {arXiv:1508.01798}
  \BibitemShut {NoStop}%
\end{thebibliography}%

\end{document}